\documentclass[reprint,amsmath,amssymb,aps,superscriptaddress]{revtex4-1}

\usepackage{graphicx}
\usepackage{hyperref}
\usepackage{xcolor}

\begin{document}

\title{Multimode Metrology via Scattershot Sampling}

\author{Joshua J. Guanzon}
\email{joshua.guanzon@uq.net.au}
\affiliation{Centre for Quantum Computation and Communication Technology, School of Mathematics and Physics, The University of Queensland, St Lucia, Queensland 4072, Australia}

\author{Austin P. Lund}
\affiliation{Centre for Quantum Computation and Communication Technology, School of Mathematics and Physics, The University of Queensland, St Lucia, Queensland 4072, Australia}
\affiliation{Dahlem Center for Complex Quantum Systems, Freie Universit\"at Berlin, 14195 Berlin, Germany}

\author{Timothy C. Ralph}
\affiliation{Centre for Quantum Computation and Communication Technology, School of Mathematics and Physics, The University of Queensland, St Lucia, Queensland 4072, Australia}

\date{\today}

\begin{abstract}
Scattershot photon sources are known to have useful properties for optical quantum computing and boson sampling purposes, in particular for scaling to large numbers of photons. This paper investigates the application of these scattershot sources towards the metrological task of estimating an unknown phase shift. In this regard, we introduce three different scalable multimode interferometers, and quantify their quantum Fisher information performance using scattershot sources with arbitrary system sizes. We show that two of the interferometers need the probing photons to be in certain input configurations to beat the classical shot-noise precision limit, while the remaining interferometer has the necessary symmetry which allows it to always beat the classical limit no matter the input configuration. However, we can prove all three interferometers gives the same amount of information on average, which can be shown to beat the classical precision limit. We also perform Monte Carlo simulations to compare the interferometers in different experimentally relevant regimes, as a function of the number of samples.  
\end{abstract}

\maketitle

\section{Introduction}

The science of measurement, or metrology, is the study of measurement accuracy and precision in a broad range of experimental contexts. Quantum metrology considers these factors in situations which can only be described by quantum mechanics. Measurement probes that exhibit non-classical properties, such as quantum entanglement, can in principle have advantages in measurement precision~\cite{giovannetti2006quantum,giovannetti2011advances}. In terms of optics, non-classical states of light can be used to estimate system parameters (e.g. phase shifts) with greater precision than can be achieved classically, given the same amount of probe energy~\cite{bollinger1996optimal,lee2002quantum,joo2011quantum}. This could be useful for many situations, for example, if a sample is highly photosensitive, or the number of probing measurements is limited. There is an enormous range of possibilities to consider and finding the best architecture, with an easy to create probe state, is a problem of continued research and interest. 

The concept of non-classicality is not one that has a universal definition across the fields of study based on quantum mechanics.  In the field of quantum computing, a quantum resource that can be used to exhibit a computational speed-up of the same problem using classical resources would be considered ``non-classical.''  In optical quantum computing, the surprisingly simple system of passive linear optical networks with multiple single photon inputs and detection exhibits such an advantage~\cite{tillmann2013experimental,spring2013boson,broome2013photonic}. This task is called boson sampling, where it is computationally difficult for classical computers to approximately reproduce samples from a randomly chosen passive linear optical network~\cite{aaronson2011computational}; this is due to the quantum photon number-path entanglement. The latest experimental boson sampling type devices have now reached system sizes where the quantum computational advantage is overwhelming in comparison to the state-of-the-art classical simulation strategies and supercomputers~\cite{zhong2020quantum}. 

Evidently, it would be fruitful to consider whether boson sampling-like systems are useful for metrology. In this regard, there has been recent papers which studied a multimode metrology scheme based on Quantum Fourier Transformation (QFT) interferometers, which shows a quantum advantage up to certain network sizes~\cite{motes2015linear,olson2017linear}. Like boson sampling, these devices use single photon inputs where the QFT induces number-path entanglement to beat the classical precision limit. This scheme with QFT interferometers has been implemented experimentally in various small sizes~\cite{su2017multiphoton}. 

\begin{figure}[htbp]
    \begin{center}
        \includegraphics[width=\linewidth]{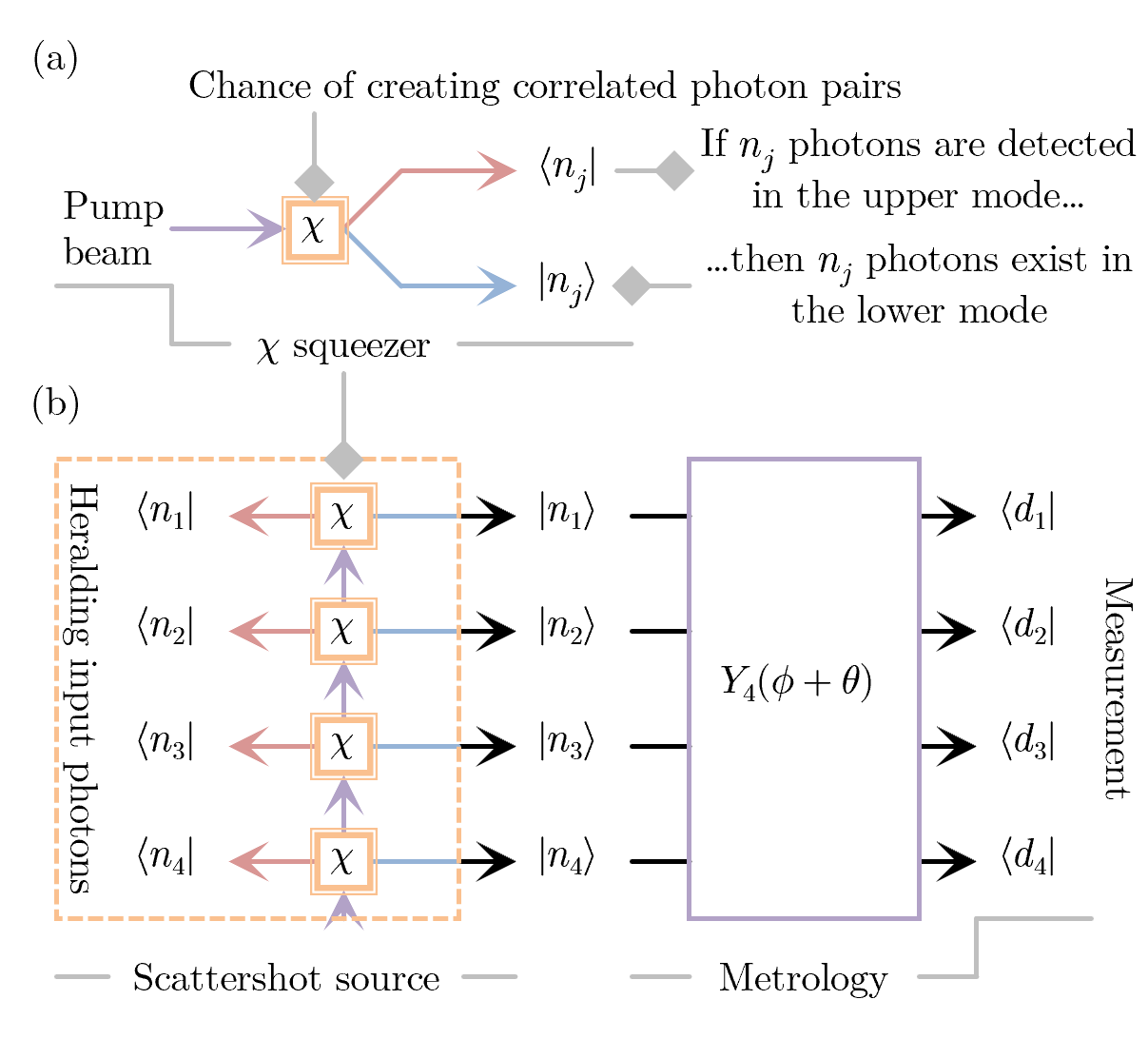}
        \caption{\label{fig:scattershot} 
            (a) A two-mode squeezer, characterised by a $\chi$ squeezing parameter, has a chance of creating pairs of photons via spontaneous parametric down-conversion of pump photons. This means if we detect $n_i$ photons in the red upper mode, we expect to find $n_i$ photons in the blue lower mode. 
            (b) By stacking an $m$ array of $\chi$ strength squeezers and heralding photo-detectors (here for $m=4$), we can create a scattershot source which allows us to quickly perform metrology experiments with good potential for scaling to higher modes.}
    \end{center}
\end{figure}

\begin{figure*}[htbp]
    \begin{center}
        \includegraphics[width=\linewidth]{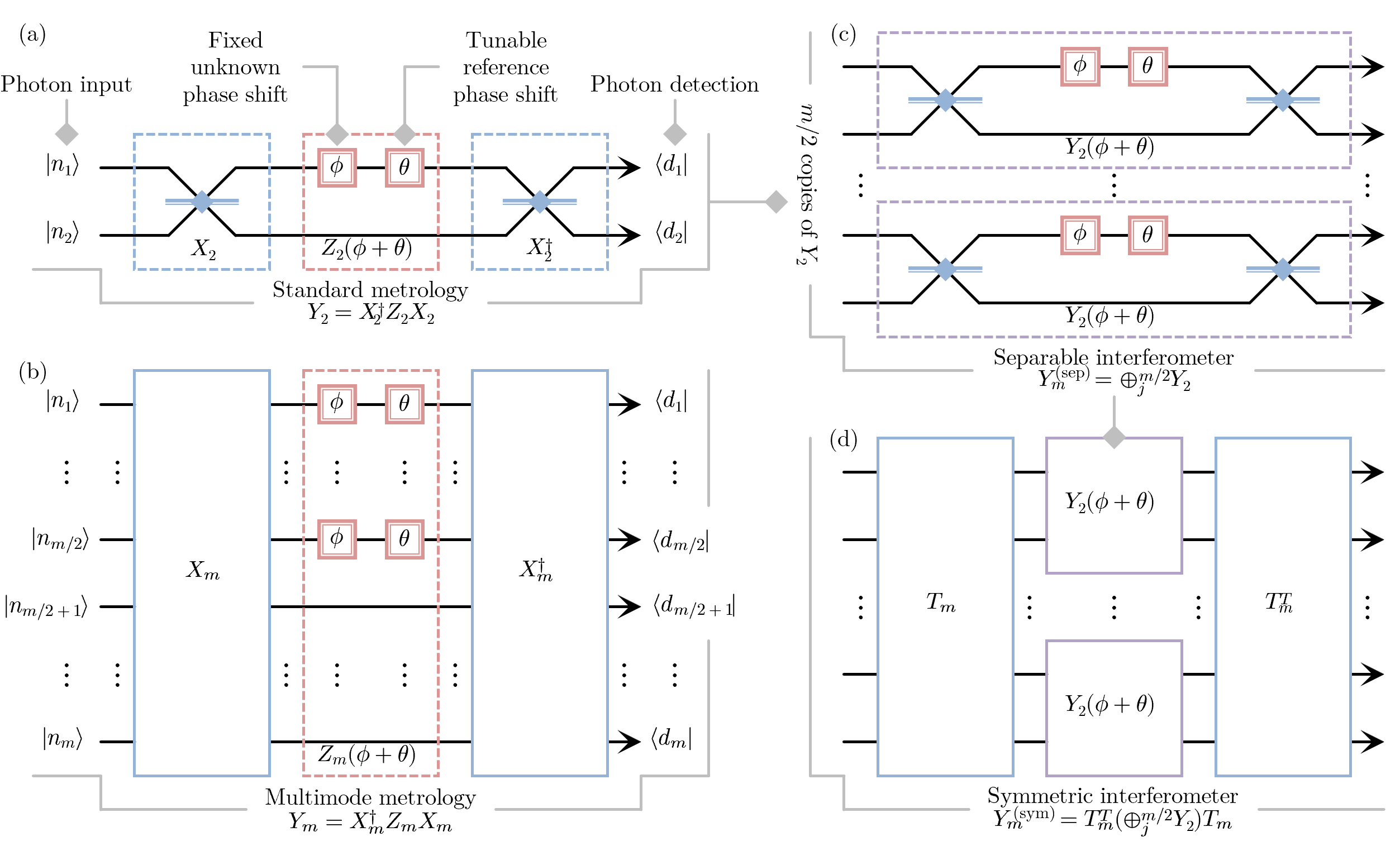}
        \caption{\label{fig:metrology} 
            (a) A standard metrology experiment to determine $\phi$, using a Mach–Zehnder interferometer, photon inputs and photon counting detectors. 
            (b) We explore the natural multimode version of this standard setup, where we investigate interferometers with phase shifts applied to half of the $m$ modes. 
            (c) One multimode interferometer of this type is made from stacking together $m/2$ copies of Mach-Zehnder interferometers. However, the amount of information gathered about $\phi$ depends on which ports the photons are injected into.
            (d) We will show that we can create a symmetrising transformation $T_m$, which results in all mode being treated equally, just like in the standard two-mode metrology experiment.
            } 
    \end{center}
\end{figure*}

The technical demands on the photonic sources required for boson sampling experiments was eased with the development of \textit{scattershot} boson sampling. This removes the strict resource requirement of single photon inputs into particular modes, and instead allowed random configurations of photonic inputs~\cite{lund2014boson,bentivegna2015experimental,zhong201812}. Scattershot boson sampling is still a computationally hard problem for classical computers, while being efficiently implementable on a quantum optical device that employs a source with much improved scalability to higher modes compared to regular boson sampling using single photons. The main building block of this source is a standard two-mode squeezed vacuum with a heralding photon detector in one of the modes, as shown in Fig.~\ref{fig:scattershot}(a). The scattershot source is then built from an array of these components as shown in Fig.~\ref{fig:scattershot}(b), whose photonic emissions are then fed into a boson sampling circuit for quantum computation purposes, or, as in our case, fed into a multimode interferometer for quantum metrology purposes. 

In essence, similar to how Ref.~\cite{motes2015linear} translated the ideas from boson sampling for multimode metrology purposes, this paper aims to translate the ideas from scattershot boson sampling for scattershot multimode metrology purposes. We note that there has been some numerical analysis of scattershot sources for metrology in one section of Ref.~\cite{you2017multiparameter}, in the context of multi-parameter estimation with QFT interferometers. In contrast, our paper has scattershot sources as the main stage of our single-parameter estimation scheme, with different phase shift scaling considerations, and analysis of different multimode interferometers beyond that of QFTs. 

We begin in Section~\ref{sec:char}, where we will outline the structure of the metrology interferometers which we are investigating. We also give background information on passive linear optical networks and the Quantum Fisher Information (QFI). In Section~\ref{sec:sep}, we analyse the separable interferometer, which we will argue is the most natural baseline to compare other multimode interferometers of similar structure. We then show in Section~\ref{sec:uni} that the uniform interferometer, which uses QFTs in the manner like that to the previously mentioned papers, actually doesn't lead to the required symmetry that maximises the benefits of using scattershot sources. In Section~\ref{sec:sym}, we outline how to construct the symmetric interferometer, and prove that the overall network symmetry means that all possible photon input configurations lead to a quantum enhancement from scattershot sources. We calculate the QFI for all three interferometers, which shows all three can beat the classical shot-noise precision limit, and in the asymptotic limit of large numbers of samples all three will give the same amount of information. Finally, in Section~\ref{sec:advan}, we implement Monte Carlo simulations to contrast the interferometers with finite sample sizes, and in particular regimes that may be of experimental interest.

\section{Characterising the Metrology Apparatus} \label{sec:char}

\subsection{Passive Linear Optical Networks} 

The way a particular linear optical network evolves photonic inputs can be described simply by an $m \times m$ unitary matrix $U_m$, where $m$ is the number of modes or optical ports that the network contains. More precisely, this matrix describes how the bosonic creation operators of the input modes $\vec{a}^\dag$ are linearly combined to give the bosonic creation operators of the output modes 
\begin{align}
    \vec{b}^\dag = U_m\vec{a}^\dag.
\end{align}
In other words, $U_m$ describes how the photons are scattered and interfered within the network, such that the number of photons and energy are conserved (assuming a lossless network). 

There are various convenient methods of decomposing any unitary $U_m$ into a maximum of $m(m-1)/2$ elementary beam-splitters \cite{reck1994experimental,nielsen2010quantum,clements2016optimal}. Hence linear optical networks are accessible systems to study, as they can be implemented experimentally using just conventional one- and two-mode linear optical elements. We are also not considering networks with internal active components, which is where inter-network measurements are taken to change the configuration of other optical elements elsewhere within the network. This makes the passive networks we describe in this paper amenable to implementation on miniaturised optical interferometers and integrated optical circuits in the near-term.

\subsection{Multimode Interferometers} 

In a metrology experiment, we want to maximise our knowledge of a fixed unknown phase shift $\phi$, while minimising the amount of resources it takes to get that information. First, consider the standard two-mode Mach–Zehnder interferometer (MZI) given by the unitary
\begin{align}
    Y_2(\phi) \equiv X_2 Z_2(\phi) X_2^\dagger =         
        \begin{bmatrix}
            \cos(\phi/2) & \sin(\phi/2) \\ 
            -\sin(\phi/2) & \cos(\phi/2) \\
        \end{bmatrix}, \label{eq:Y2}
\end{align}
where the unknown phase shift $Z_2(\phi)$ is applied to only one of the modes, and is conjugated by fixed 50:50 symmetric beam-splitters $X_2$. This is shown visually in Fig.~\ref{fig:metrology}(a), where we have included a controllable known phase shift $\theta$~\footnote{Whilst the phase parameter $\theta$ is variable, it is not considered to be changed whilst photons are evolving through the network.  Rather, it is considered to be set and fixed before samples are recorded.  Therefore the presence of this controllable phase shift is not in contradiction to our ``passive network'' premise.}, as a reminder that the overall phase shift could always be tuned close to an experimentally convenient value (e.g. $\phi+\theta\approx0$ such that the overall scattering is close to the identity $Y_2(0)=\mathbb{I}_2$). However, for mathematical convenience, we will set $\theta=0$ and just remember we can always tune the phase shift. 

Now, consider the $m$-mode extension of this interferometer
\begin{align}
    Y_m(\phi) \equiv X_m Z_m(\phi) X_m^\dagger, 
\end{align}
where the phase shift $Z_m(\phi)$ is over half of the modes, and $X_m$ is a fixed linear optical network, analogous to the 50:50 beam-splitters in the two-mode MZI, but $X_m$ could potentially mix between all modes. This arrangement is shown in Fig.~\ref{fig:metrology}(b).  Using this network construction we will explore different $X_m$ networks which will allows us to extend particular properties of the standard MZI to multiple modes. It is important to note that the treatment of the phase shift $Z_m(\phi)$ as a resource is different to the previously mentioned Ref.~\cite{olson2017linear}, where they consider the best way to divide up a fixed total amount of phase amongst many modes (i.e. the phase shift can be different on each mode depending on the strategy employed). In our case, the phase shift applied on the modes is the same; physically, we are just considering a straight-forward scenario where the sample being measured, such as a piece of glass or ampule of gas, is uniform and simply overlaid upon half of the modes. 

One benefit of investigating interferometers of this type is that a separable stack of $m/2$ MZIs  
\begin{align}
    Y_m^{(\text{sep})}(\phi) \equiv \oplus^{m/2}_{j=1} Y_2(\phi),
\end{align}
is an interferometer that fits the phase structure as shown in Fig.~\ref{fig:metrology}(c), where effectively $X_m^{(\text{sep})}=\oplus^{m/2}_{j=1}X_2$ with some trivial rearrangement of the modes. It is known that photon inputs into a MZI can show quantum enhanced detection, as long as the photons are not all in one port~\cite{takeoka2017fundamental}; hence scaling this system by $m/2$ copies effectively expedites determining $\phi$. Alternatively, $Y_m^{(\text{sep})}(\phi)$ can be thought of as being separated temporally, in other words just using one standard MZI with $m/2$ samples. Given independent samples, the total information will be just the sum of the individual measurements; the precise QFI is calculated in Section~\ref{sec:sep}. It is this intuitive interpretation which we use to fairly compare different sized multimode interferometers with the same structure, as we can use separable interferometers with equivalent number of phase shifts as a standard metric. 

For our metrology considerations, the two input ports of an MZI are permutationally invariant, in the sense that the inputs $|n_1n_2\rangle$ and $|n_2n_1\rangle$ will provide the same amount of extractable information about $\phi$; this is the natural consequence of the 50:50 beam splitters $X_2$. In our multimode case, let us suppose the scattershot source gave a particular input of $| \vec{n} \rangle \equiv |n_1 \cdots n_m \rangle$, input invariance would mean the QFI remains the same if we switch any two modes $n_j \leftrightarrow n_k,\forall j,k$. Since the scattershot source generates photons at random inputs with equal chance in all modes, it would be advantageous if we find a $Y_m(\phi)$ interferometer which has this input invariance property for all modes. Clearly for $Y_m^{(\text{sep})}(\phi)$ the first mode interacts with the second mode, but doesn't interact with the third mode, hence it does not have input invariance. One may speculate that this property exists for 
\begin{align}
    Y_m^{(\text{uni})}(\phi) \equiv F_m Z_m(\phi) F_m^\dagger, 
\end{align}
where we replace $X_m^{(\text{uni})}=F_m$ in Fig.~\ref{fig:metrology}(b) with a uniform scattering device such as the QFT or Hadamard (Sylvester) transformation. These are optical devices in which the associated scattering matrix has elements of the same uniform magnitude, which means if a single photon is injected into any of the input ports, there is an equal chance of measuring it in any of the output ports. However, we will show in Section~\ref{sec:uni} that uniform interferometers do not result in the symmetry which we are looking for; hence finding an interferometer with input invariance is not an immediately obvious problem to solve. 

We propose using a symmetrising transformation $T_m$, which is used to create the interferometer 
\begin{align}
    Y_m^{(\text{sym})}(\phi) &\equiv T_m [\oplus^{m/2}_i Y_2(\phi)] T_m^T, \label{eq:symYm}
\end{align}
that has a symmetrical scattering matrix with very similar properties as $Y_2(\phi)$. This construction is shown in Fig.~\ref{fig:metrology}(d), where the symmetric interferometer clearly has the same overall phase structure as the other interferometers, however with $X_m^{(\text{sym})} = T_m(\oplus^{m/2}_{j=1}X_2)$ and some trivial rearrangement of the modes. We detail this in Section~\ref{sec:sym}, and show that this is an optimal way of mixing the modes such that it is input invariant. In other words, we can show that the symmetric interferometer will \emph{always} give a quantum enhanced detection using scattershot sources. As a consequence of this, we show in Section~\ref{sec:advan} that $Y_m^{(\text{sym})}(\phi)$ is more likely to give more information than $Y_m^{(\text{sep})}(\phi)$ for finite sample sizes. Finally, we note that the symmetric interferometer can actually be implemented with temporal modes and one MZI, where $T_m$ describes how the different temporal modes should interact together; hence the experimental implementation of this scheme to large mode numbers is feasible. 

\subsection{Quantum Fisher Information}

The Cramer-Rao bound constrains the achievable precision of an unknown variable $\phi$ as follows
\begin{align}
    (\Delta \phi)^2 \geq \frac{1}{\mathcal{F}}, 
\end{align}
where $\mathcal{F}$ is the QFI~\cite{helstrom1976quantum}. Classical interferometers cannot beat the shot-noise limit (SNL) of $(\Delta \phi)^2 \geq 1/n$, where $n$ here is the total number of probes or photons in our case. In contrast, it is known that quantum interferometers can achieve the higher precision Heisenberg limit of $(\Delta \phi)^2 \geq 1/n^2$, through quantum entangled photons or squeezed states~\cite{escher2011general,demkowicz2012elusive}. 

Suppose we know that an interferometer is described by the unitary operator $\hat{U}(\phi)$, where an input state $\rho$ is acted upon as follows
\begin{align}
    \rho(\phi)=\hat{U}(\phi)\rho \hat{U}^\dagger(\phi),\quad \hat{U}(\phi)=e^{-i\hat{H}\phi},
\end{align}
in which $\hat{H}$ is the generating Hermitian operator for $\hat{U}$. We can then use the following equation to calculate the QFI 
\begin{align}
    \mathcal{F} = 4(\Delta \hat{H})^2 = 4(\langle\vec{n}|\hat{H}^2|\vec{n}\rangle-\langle\vec{n}|\hat{H}|\vec{n}\rangle^2), \label{eq:QFI}
\end{align}
since our input from the scattershot source is a pure state~\cite{braunstein1994statistical}, as a tensor product of Fock states 
\begin{align}
    |\vec{n}\rangle \equiv |n_1\rangle \otimes \cdots \otimes |n_m\rangle \equiv |n_1\cdots n_m\rangle. 
\end{align}
This computation is convenient as it allows us to determine the QFI just from the input state, seemingly independent of the detection scheme at the output. However, since it in some sense optimises over all possible measurement schemes, it may include schemes which requires prior knowledge of $\phi$. Therefore, in order to be certain that this calculation is realistic, we will also describe a specific measurement procedure which can independently be shown to achieve the same QFI, and hence the same precision bound on $\phi$.

\section{The Separable Interferometer} \label{sec:sep}

\subsection{General Fisher Information}

\begin{table*}[htbp]
    \caption{\label{tab:table2}
        Summary of the Quantum Fisher Information associated with the three multimode interferometers under investigation. We also include the example of two single photons into four modes $m=4$, which emphasises the input invariance of the symmetric interferometer, and that on average all three will give the same amount of information. 
    }
    \begin{ruledtabular}
        \begin{tabular}{ c c c c c c c c c }
            Input & $|\vec{n}\rangle \equiv |n_1 \cdots n_m\rangle $ & $|1100\rangle$ & $|1010\rangle$ & $|1001\rangle$ & $|0110\rangle$ & $|0101\rangle$ & $|0011\rangle$ & Avg. \\
            \hline
            Separable & $n + 2\sum_{j=1}^{m/2}n_{2j-1}n_{2j}$ & 4 & 2 & 2 & 2 & 2 & 4 & 8/3 \\ 
            Uniform & $n + \frac{8}{m^2}\sum_{j=1}^{m/2} \sum_{k=1}^{m/2} n_{2j-1}n_{2k}/\sin^{2}\left(\frac{\pi(2k-2j+1)}{m}\right)$ & 3 & 2 & 3 & 3 & 2 & 3 & 8/3 \\
            Symmetric & $n + \frac{1}{m-1}\sum_{j=1}^m \sum_{k \neq j}^m n_j n_k$ & 8/3 & 8/3 & 8/3 & 8/3 & 8/3 & 8/3 & 8/3
        \end{tabular}
    \end{ruledtabular}
\end{table*}

The Hermitian operator which describes a single MZI is given by $\hat{H}^{(\text{mzi})}= -\frac{i}{2}(a_1^\dagger a_2 - a_2^\dagger a_1)$, since Eq.~\eqref{eq:Y2} is just the beam-splitter action with an extra $1/2$. Thus a separable system of $m/2$ MZIs performs the action 
\begin{align}
    \hat{H}^{(\text{sep})} = -\frac{i}{2}\sum_{j=1}^{m/2}(a_{2j-1}^\dagger a_{2j} - a_{2j}^\dagger a_{2j-1}). 
\end{align}
To calculate the QFI using Eq.~\eqref{eq:QFI}, we need an expression for $\langle\vec{n}|\hat{H}^{(\text{sep})}|\vec{n}\rangle$ and $\langle\vec{n}|(\hat{H}^{(\text{sep})})^2|\vec{n}\rangle$. All the terms in $\hat{H}^{(\text{sep})}$ will not preserve $|\vec{n}\rangle$ and thus will give zero, for example $\langle n_1 n_2 | a_1^\dagger a_2 | n_1 n_2 \rangle \propto \langle n_1 n_2 | (n_1+1) (n_2-1) \rangle = 0$, hence $\langle \vec{n}|\hat{H}^{(\text{sep})}|\vec{n}\rangle = 0$. In contrast, the square of this Hermitian operator
\begin{align}
    (\hat{H}^{(\text{sep})})^2 &= \frac{1}{4}\sum_{j=1}^{m/2}(a_{2j-1}^\dagger a_{2j-1} a_{2j}a_{2j}^\dagger \nonumber \\ 
    &\quad + a_{2j-1}a_{2j-1}^\dagger a_{2j}^\dagger a_{2j}) \nonumber \\
    &\quad + \text{non-state-preserving terms},
\end{align}
contains important state preserving terms which contribute to the QFI, such as $\langle n_1 n_2 | a_1^\dagger a_1 a_2 a_2^\dagger | n_1 n_2 \rangle = n_1(n_2+1)$. Thus, this means that 
\begin{align}
    \langle\vec{n}|(\hat{H}^{(\text{sep})})^2|\vec{n}\rangle &= \frac{1}{4}\sum_{j=1}^{m/2}[n_{2j-1}(n_{2j}+1) \nonumber \\ 
    &\quad + n_{2j}(n_{2j-1}+1)]. 
\end{align}
Now, substituting this into Eq.~\eqref{eq:QFI} results in a QFI of 
\begin{align}
    \mathcal{F}^{(\text{sep})} &= \sum_{j=1}^{m/2}[n_{2j-1}+n_{2j}+2n_{2j-1}n_{2j}], \nonumber \\ 
    &= n + 2\sum_{j=1}^{m/2}n_{2j-1}n_{2j}, \label{eq:sepF}
\end{align}
where we define $n=\sum_{j=1}^m n_j$ as the total number of photons. So this means that this separable interferometer beats the shot-noise limit of $n$, as long as some of the photons land together in one of the MZIs. We note that this QFI calculation matches the results from Ref.~\cite{perarnau2020multimode}, which analysed an analogous system of a multimode, temporal-spaced Fock input state injected into a single MZI. This only emphasises how we can interpret this result as effectively repeating a single standard MZI experiment $m/2$ times, and hence is the natural baseline to compare with other multimode interferometers of similar phase structure. Evidently, for $m=2$ we can see this result reduces down to
\begin{align}
    \mathcal{F}^{(\text{mzi})} = \mathcal{F}^{(\text{sep})}(m=2) = n_1 + n_2 + 2n_1n_2, 
\end{align}
the QFI for arbitrary Fock states into a single MZI, which matches known results from similar scenarios~\cite{lang2014optimal,yu2018maximal}. Note that the QFI for all multimode interferometers we consider are summarised in Table~\ref{tab:table2}. 

\subsection{Specific Measurement Procedure}

In this section, we detail a concrete measurement procedure, in particular for single photon input states $|\vec{n}\rangle\equiv|n_1\cdots n_m\rangle, n_i\in\{0,1\}$, which saturates the previous $\mathcal{F}^{(\text{sep})}$ result. The Fisher information from single photons into an individual MZI $Y_2$ is given by  
\begin{align}
    \mathcal{F}^{(\text{mzi})}(|10\rangle) &= \mathcal{F}^{(\text{mzi})}(|01\rangle) = 1, \\ 
    \mathcal{F}^{(\text{mzi})}(|11\rangle) &= 4. \label{eq:mzi11}
\end{align}
We get Eq.~\eqref{eq:mzi11} from previous known results of Holland-Burnett states~\cite{holland1993interferometric,datta2011quantum}, in which Ref.~\cite{datta2011quantum} in particular describes a measurement procedure with photon number resolving detectors. We can use this knowledge and counting to determine the Fisher information received from $n$ single photons randomly allocated into the $m$ ports of $Y_m^{(\text{sep})}\equiv\oplus_{j=1}^{m/2}Y_2$. Suppose that out of $n$ separate photons, there are $x$ pairs which land together into one sub-experiment as $Y_2|11\rangle$. This translates to a QFI of the following
\begin{align}
    \mathcal{F}^{(\text{sep})}(|\vec{n}\rangle,x) &= x \mathcal{F}^{(\text{mzi})}(|11\rangle) + (n-2 x)\mathcal{F}^{(\text{mzi})}(|10\rangle), \nonumber \\ 
    &= n + 2x,\quad n_i\in\{0,1\}. 
\end{align}
This result matches the general QFI calculation in Eq.~\eqref{eq:sepF}, where $x=\sum_{j=1}^{m/2}n_{2j-1}n_{2j}$ for single photons. 

Since the scattershot source treats all modes equally, there is an equal chance of generating any of the $\binom{m}{n}$ cases. We show in Appendix~\ref{sec:sepcomb} that the probability $x$ pairs occurs is 
\begin{align}
    \mathcal{P}(x) = \binom{m}{n}^{-1}\binom{m/2}{x} \binom{m/2-x}{n-2x} 2^{n-2x}, \label{eq:sepPx}
\end{align}
where $\sum_{x=0}^{n/2} \mathcal{P}(x) = 1$. We can then calculate the expected Fisher information for our separable circuit with single photons as 
\begin{align}
    \langle \mathcal{F}^{(\text{sep})} \rangle = \sum_{x=0}^{n/2} \mathcal{F}^{(\text{sep})}(x) \mathcal{P}(x) = n + \frac{n(n-1)}{m-1}, \label{eq:sepFavg}
\end{align}
hence on average we expect that the separable interferometer coupled with a scattershot source will provide a quantum advantage. We will show in Section~\ref{sec:sym} that this expression is equivalent to the QFI of the symmetric interferometer $\mathcal{F}^{(\text{sym})}$, if we also assume single photon inputs.

\section{The Uniform Interferometer} \label{sec:uni}

Here we calculate the QFI using uniform scattering devices, and show that it isn't input invariant. Firstly, note that if we replace $X_m$ with a Hadamard transform, we get back the separable circuit but with the modes switched around; so instead we will focus on the non-trivial QFT based interferometer $X_m^{(\text{uni})}=F_m$. We show in Appendix~\ref{sec:uniherm} the scattering matrix which describes the overall uniform interferometer is
\begin{align}
    (Y^{(\text{uni})}_m)_{j,k} &= \begin{cases}
        \cos(\phi/2), & j = k, \\
        0, & k - j \in \text{even}/\{0\}, \\
        \frac{ 4i \sin(\phi/2) }{m\left(1-e^{\frac{2i\pi(k-j)}{m}}\right)}, & k - j \in \text{odd},
    \end{cases} \label{eq:uniYm}
\end{align}
and the corresponding Hermitian operator is 
\begin{align}
    \hat{H}^{(\text{uni})} = \frac{2}{m} \sum^{m/2}_{j=1} \sum^{m/2}_{k=1} \left[ \tfrac{a^\dagger_{2j-1}a_{2k}}{1-e^{\frac{2i\pi(2k-2j+1)}{m}}} + \tfrac{a^\dagger_{2k}a_{2j-1}}{1-e^{-\frac{2i\pi(2k-2j+1)}{m}}} \right]. 
\end{align}
Firstly note that $\langle\vec{n}| \hat{H}^{(\text{uni})}|\vec{n}\rangle = 0$ because there are no state preserving terms. The Hermitian operator squared is given by 
\begin{align}
    (\hat{H}^{(\text{uni})})^2 &= \frac{1}{m^2} \sum_{j=1}^{m/2} \sum_{k=1}^{m/2} \tfrac{a^\dagger_{2j-1}a_{2j-1}a_{2k}a_{2k}^\dagger + a_{2j-1} a^\dagger_{2j-1}a_{2k}^\dagger a_{2k}}{\sin^2\left(\frac{\pi(2k-2j+1)}{m}\right)} \nonumber \\ 
    &\quad + \text{non-state-preserving terms}.
\end{align}
Hence we can calculate the expected value as 
\begin{align}
    \langle\vec{n}|(\hat{H}^{(\text{uni})})^2|\vec{n}\rangle &= \frac{1}{m^2} \sum_{j=1}^{m/2} \sum_{k=1}^{m/2} \tfrac{n_{2j-1}(n_{2k}+1) + n_{2k}(n_{2j-1}+1)}{\sin^2\left(\frac{\pi(2k-2j+1)}{m}\right)}. 
\end{align}
There is a simplified expression for this type of cosecant summation~\cite{cosecant}, which we use to show that $\sum_{k=1}^{m/2} \tfrac{1}{\sin^2\left(\frac{\pi(2k-2j+1)}{m}\right)} = m^2/4$. Finally, the QFI is 
\begin{align}
    \mathcal{F}^{(\text{uni})} &= \frac{4}{m^2} \sum_{j=1}^{m/2} \sum_{k=1}^{m/2} \tfrac{n_{2j-1}+n_{2k}+2n_{2j-1}n_{2k}}{\sin^2\left(\frac{\pi(2k-2j+1)}{m}\right)}, \nonumber \\ 
    &= \sum^{m/2}_{j=1} n_{2j-1} + \sum^{m/2}_{k=1} n_{2k} \nonumber \\ &\quad + \frac{8}{m^2} \sum_{j=1}^{m/2} \sum_{k=1}^{m/2} \tfrac{n_{2j-1}n_{2k}}{\sin^2\left(\frac{\pi(2k-2j+1)}{m}\right)}, \nonumber \\
    &= n + \frac{8}{m^2} \sum_{j=1}^{m/2} \sum_{k=1}^{m/2} \tfrac{n_{2j-1}n_{2k}}{\sin^2\left(\frac{\pi(2k-2j+1)}{m}\right)}. \label{eq:uniF}
\end{align}
We can see that this interferometer can beat shot-noise, however the input must have at least one photon in an odd mode and at least one photon in an even mode. In other words, if the scattershot source happens to generate photons which are all in the odd modes, then the amount of information will just be equivalent to shot-noise (and likewise if they are all in the even modes). 

We can gain a good sense of the possible magnitudes of the coefficients in front of the enhancement terms in Eq.~\eqref{eq:uniF}, by noting that the denominator is smallest when $k-j=0$ and that $\sin^2\left(\pi/m\right)\approx \pi^2/m^2$ assuming our system has modes $m \gg \pi$. Therefore, the coefficients should be between $(8/m^2, 8/\pi^2 \approx 0.81)$, clearly smaller than the fixed $2$ coefficient in front of the enhancement terms for the separable interferometer. However, considering the scattershot source gives a random input, this means that the uniform interferometer has a much higher probability of beating the shot-noise limit. In the next section, we will detail the existence of a symmetric interferometer that will always beat the shot-noise limit, no matter the location of the photons.

\section{The Symmetric Interferometer} \label{sec:sym}

\subsection{Creating the Symmetrising Transformation} 

Consider a metrology apparatus $Y_m^{(\text{sym})}$ which gives the same amount of information about $\phi$ irrespective of where we probe it with photons; intuitively, we expect there to be symmetry associated with how the photons are scattered within $Y_m^{(\text{sym})}$. It will be shown that the necessary network symmetry is encapsulated by a type of skewed-symmetric matrix, that was used in a different context in Ref.~\cite{guanzon2020controllable}. But first, we will describe one method of creating the symmetrising transformation $T_m$, which acts on $\oplus^{m/2}_{j=1} Y_2$ and gives us this overall network symmetry. 

The $m$ mode symmetrising transformation $T_m$ is represented by a $m\times m$ real orthogonal matrix, and can be built using two smaller $2\times2$ submatrices as follows 
\begin{align}
    T_m &\equiv 
       \overbrace{ \begin{bmatrix}
             A_2 &  B_2 &  B_2 &  B_2 & \cdots \\ 
            -B_2 &  A_2 & -B_2 &  B_2 & \cdots \\
            -B_2 &  B_2 &  A_2 & -B_2 & \cdots \\ 
            -B_2 & -B_2 &  B_2 &  A_2 & \cdots \\ 
            \vdots & \vdots & \vdots & \vdots & \ddots 
        \end{bmatrix} }^{m/2 \text{ columns}} 
        \left. \phantom{\begin{bmatrix}
             A_2 \\ 
            -B_2 \\
            -B_2 \\ 
            -B_2 \\ 
            \vdots  
        \end{bmatrix}\hspace{-3.6em}} \right\} \rotatebox[origin=c]{-90}{\scriptsize $m/2$ rows}, \label{eq:Tm} \\ 
            A_2(m) &\equiv 
        \begin{bmatrix}
             0 & \frac{1}{\sqrt{2}} \\ 
            \frac{\sqrt{m}-\sqrt{(m-1)(m-2)}}{\sqrt{2m(m-1)}} & -\sqrt{\frac{m-2}{2m(m-1)}} 
        \end{bmatrix}, \label{eq:A2} \\
    B_2(m) &\equiv
        \begin{bmatrix}
             \frac{1}{\sqrt{2(m-1)}} & \sqrt{\frac{m}{2(m-1)(m-2)}} \\ 
            \frac{\sqrt{m(m-2)}+2\sqrt{m-1}}{\sqrt{2m(m-1)(m-2)}} & -\sqrt{\frac{m-2}{2m(m-1)}} 
        \end{bmatrix}. \label{eq:B2}
\end{align}
Notice in Eq.~\eqref{eq:Tm} that some of the off-diagonal entries are negative (i.e. $-B_2$). We purposefully set these negative entries such that $T_m$ has lower-resolution orthogonality and off-diagonal skew-symmetry in $2\times2$ blocks. We can use other orthogonal and skew-symmetric matrices to help us determine which entries need to be negative. Specifically in our case, we can construct the following helper matrices $H_{2n}$, for all powers of two sizes $\forall n=2^k, k\in \mathbb{N}$ as follows 
\begin{align}
    H_{2n} &\equiv \frac{\sqrt{n-1}}{\sqrt{2n-1}}
        \begin{bmatrix}
            H_n & H_n + \frac{\mathbb{I}_n}{\sqrt{n-1}} \\ 
            H_n - \frac{\mathbb{I}_n}{\sqrt{n-1}} & -H_n \\
        \end{bmatrix}, \label{eq:H2n}
\end{align}
in which the first two sizes will be 
\begin{align}
    H_2 &\equiv 
        \begin{bmatrix}
            0 & 1 \\ 
            -1 & 0 \\
        \end{bmatrix} 
    \Rightarrow H_4 = \frac{1}{\sqrt{3}}\begin{bmatrix}
             0 &  1 &  1 &  1 \\ 
            -1 &  0 & -1 &  1 \\
            -1 &  1 &  0 & -1 \\ 
            -1 & -1 &  1 &  0 
        \end{bmatrix}. 
\end{align}
Notice that we used the location of the negative entries in $H_4$ for the first few entries of $T_m$ in Eq.~\eqref{eq:Tm}. By construction, these $H_{2n}$ matrices are orthogonal $H_{2n}H_{2n}^T=\mathbb{I}_{2n}$, skew-symmetric $H_{2n}^T=-H_{2n}$ and have equal magnitude off-diagonal elements $|(H_{2n})_{i,j}|=1/\sqrt{2n-1}, \forall i \neq j$~\cite{guanzon2020controllable}. Note that $H_n$ is effectively a modified skew Hadamard matrix, and it's possible to instead use the position of the negatives from any other $m/2$-sized skew Hadamard matrix to create $T_m$. It is roughly a century old conjecture in mathematics that Hadamard matrices exist for $1$, $2$ and $4k$ sizes~\cite{hedayat1978hadamard}. Hence it is reasonable to assume that these $T_m$ transformations are possible only for $m\in\{4,8k\}$ modes, in which we showed by $H_{2n}$ an explicit method to construct these matrices for powers of two. 

Due to the above defined properties of $T_m$, we can calculate the following
\begin{align}
    T_m T_m^T &= 
        \begin{bmatrix}
             \mathbb{I}_2 &  0_2 &  0_2 &  0_2 & \cdots \\ 
            -0_2 &  \mathbb{I}_2 & -0_2 &  0_2 & \cdots \\
            -0_2 &  0_2 &  \mathbb{I}_2 & -0_2 & \cdots \\ 
            -0_2 & -0_2 &  0_2 &  \mathbb{I}_2 & \cdots \\ 
            \vdots & \vdots & \vdots & \vdots & \ddots 
        \end{bmatrix}, \\
    \mathbb{I}_2 &= A_2 A_2^T + (m/2-1) B_2 B_2^T, \label{eq:I2} \\
    0_2 &= B_2 A_2^T - A_2 B_2^T. \label{eq:02}
\end{align}
In particular, notice in Eq.~\eqref{eq:02} that all but two of the terms cancelled each other out due to the imposed negatives and $2\times2$ block orthogonality condition. By direct substitution of the definitions of $A_2$ and $B_2$, we can show that in fact Eq.~\eqref{eq:I2} and Eq.~\eqref{eq:02} resolves to the identity and zero matrix, respectively. Hence $T_mT_m^T = \mathbb{I}_m$, and thus $T_m$ is actually completely orthogonal, which means it can be implemented using simple linear optics. For completeness, we decomposed the four mode $T_4$ symmetrising transformation into elementary optical elements in Appendix~\ref{sec:symdecomp}. 

Finally, we find that conjugation of an array of MZIs $\oplus^{m/2}_{j=1} Y_2$ with the symmetrising transformation $T_m$, as given in Eq.~\eqref{eq:symYm}, results in the following matrix
\begin{align}
    Y_m^{(\text{sym})} &= 
        \begin{bmatrix}
             D_2 &  G_2 &  G_2 &  G_2 & \cdots \\ 
            -G_2 &  D_2 & -G_2 &  G_2 & \cdots \\
            -G_2 &  G_2 &  D_2 & -G_2 & \cdots \\ 
            -G_2 & -G_2 &  G_2 &  D_2 & \cdots \\ 
            \vdots & \vdots & \vdots & \vdots & \ddots 
        \end{bmatrix}, \label{eq:Ym2} \\
    D_2 &= A_2 Y_2 A_2^T + \left(\tfrac{m}{2}-1\right) B_2 Y_2 B_2^T = \begin{bmatrix}
            c & -s \\ 
            s & c \\
        \end{bmatrix}, \label{eq:F2} \\
    G_2 &= B_2 Y_2 A_2^T - A_2 Y_2 B_2^T = \begin{bmatrix}
            s & s \\ 
            s & -s \\
        \end{bmatrix}, \label{eq:G2} \\
    c &= \cos (\phi/2),\quad s = \frac{\sin (\phi/2)}{\sqrt{m-1}}, 
\end{align}
where we solved $D_2$ and $G_2$ by direct substitution of Eq.~\eqref{eq:A2} and Eq.~\eqref{eq:B2}. Explicitly, we have shown the overall metrology apparatus is associated with a scattering matrix with the form 
\begin{align}
    Y_m^{(\text{sym})} &= 
       \overbrace{ \begin{bmatrix}
             c & -s &  s &  s & \cdots \\ 
             s &  c &  s & -s & \cdots \\
            -s & -s &  c & -s & \cdots \\ 
            -s &  s &  s &  c & \cdots \\ 
            \vdots & \vdots & \vdots & \vdots & \ddots 
        \end{bmatrix} }^{m \text{ columns}} 
        \left. \phantom{\begin{bmatrix}
             c \\ 
             s \\
            -s \\ 
            -s \\ 
            \vdots
        \end{bmatrix}\hspace{-3.0em}} \right\} \rotatebox[origin=c]{-90}{\scriptsize $m$ rows}. \label{eq:Ym}
\end{align}
The $2\times2$ block off-diagonal skew-symmetry from $T_m$ transferred to Eq.~\eqref{eq:Ym2}, hence it is clear by substituting $D_2$ and $G_2$ that the entire off-diagonals of $Y_m^{(\text{sym})}$ must be skew-symmetric. Note the striking similarity between the scattering matrix of this symmetric interferometer and the MZI in Eq.~\eqref{eq:Y2}, in contrast to the uniform interferometer in Eq.~\eqref{eq:uniYm}. We will use this skew-symmetry to prove the input location invariance of this metrology experiment. 

\subsection{General Fisher Information}

By taking the matrix logarithm of $Y_m^{(\text{sym})}$, the generating Hermitian matrix is given by $(H_m^{(\text{sym})})_{j,k} = -\frac{i}{2\sqrt{m-1}} \text{sgn}[(Y_m^{(\text{sym})})_{j,k}](1-\delta_{j,k})$, where $\text{sgn}[(Y_m^{(\text{sym})})_{j,k}]$ is the sign of the matrix element $(j,k)$ in $Y_m^{(\text{sym})}$. Hence the associated Hermitian operator performs the following action
\begin{align}
    \hat{H}^{(\text{sym})} = -\frac{i}{2\sqrt{m-1}} \sum^m_{j=1} \sum^m_{k \neq j} \text{sgn}[(Y_m^{(\text{sym})})_{j,k}] a_j^\dagger a_k. 
\end{align}
As with the other cases, the expectation value of this operator is zero $\langle\vec{n}|\hat{H}^{(\text{sym})}|\vec{n}\rangle=0$, as there are no state preserving terms. We can calculate the square of this operator as 
\begin{align}
    (\hat{H}^{(\text{sym})})^2 &= -\frac{1}{4(m-1)} \left[ \sum^m_{j=1} \sum^m_{k \neq j} \text{sgn}[(Y_m^{(\text{sym})})_{j,k}] a_j^\dagger a_k \right] \nonumber \\ &\quad \times \left[ \sum^m_{p=1} \sum^m_{q \neq p} \text{sgn}[(Y_m^{(\text{sym})})_{p,q}] a_p^\dagger a_q \right], \nonumber \\ 
    &= \frac{1}{4(m-1)}\sum^m_{j=1} \sum^m_{k \neq j} a_j^\dagger a_j a_k a_k^\dagger  \nonumber \\
    &\quad + \text{non-state-preserving terms}.
\end{align}
Since the state preserving terms are when $p=k$ and $q=j$, we calculated that $\text{sgn}[(Y_m^{(\text{sym})})_{j,k}]\text{sgn}[(Y_m^{(\text{sym})})_{k,j}] = -1$ because of the off-diagonal skew-symmetry properties of $Y_m^{(\text{sym})}$. The associated expectation value is then 
\begin{align}
    \langle\vec{n}|(\hat{H}^{(\text{sym})})^2|\vec{n}\rangle = \frac{1}{4(m-1)}\sum^m_{j=1} \sum^m_{k \neq j} n_j (n_k + 1),  
\end{align}
and hence the Fisher information can be summarised as 
\begin{align}
    F^{(\text{sym})} &= \frac{1}{m-1}\sum^m_{j=1} \sum^m_{k \neq j} (n_j + n_j n_k), \nonumber \\ 
    &= \frac{1}{m-1}\sum^m_{j=1} \left(n_j (m-1) + \sum^m_{k \neq j} n_j n_k\right), \nonumber \\ 
    &=  n + \frac{1}{m-1}\sum^m_{j=1} \sum^m_{k \neq j} n_j n_k. \label{eq:symF}
\end{align}
As we have shown, as long as the probing photons generated by the scattershot source are located in two or more modes, this interferometer will always beat the shot-noise limit. However, clearly the $1/(m-1)$ coefficient in front of the enhancement term is smaller than the $2$ for the separable interferometer, and is in between the approximate range $(8/m^2,8/\pi^2)$ for the uniform interferometer. In fact, the average scattershot probe into the separable and uniform interferometers will give the same information as the symmetric interferometer
\begin{align}
    F^{(\text{sym})} = \langle F^{(\text{sep})} \rangle = \langle F^{(\text{uni})} \rangle. 
\end{align}
The averages were determined by adding up the QFI from all the $m!$ equiprobabilistic permutations of $|n_1\cdots n_m\rangle$ and multiplying by the probability $1/m!$, as shown in detail in Appendix~\ref{sec:symavg}. Hence, in the long run with multiple scattershot samples, all three interferometers are expected to give the same amount of QFI. 

The double sum over all modes in Eq.~\eqref{eq:symF} means that if any of the input modes were switched, the amount of information from the symmetric interferometer will remain the same. Hence we have shown that this interferometer holds the same input invariance property that the standard MZI has, where each mode is treated equally from a metrological sense. Interestingly, in Appendix~\ref{sec:syminv} we can prove the input invariance of the four mode case $Y_4^{(\text{sym})}$ using a completely different method via similarity transformations. To reinforce this QFI result, we will now describe a specific measurement procedure using single photons which we show can saturate this bound. 

\subsection{Specific Measurement Procedure}

In this section, we will consider inputs where each mode can have up to one photon $n_i\in\{0,1\}$, which for scattershot sources corresponds to $n$ single photons being injected into random input ports of $Y_m^{(\text{sym})}$ with uniform probability. We do this as the single photon case is easier to verify mathematically. Furthermore, the single photons case is more experimentally practical in the near term, as it only needs small amounts of squeezing for the photon source and doesn't require number resolving detectors for the measurement. 

We consider a measurement procedure where we are tuning $\theta$ to approximately $-\phi$ (recall from Fig.~\ref{fig:metrology} that $\theta$ is a known reference phase), which turns the entire interferometer into approximately the identity $Y_m^{(\text{sym})}(\theta+\phi=0)=\mathbb{I}_m$. For simplicity, we will again ignore $\theta$ and assume $\phi \approx 0$. We can identify how close $\phi$ is to $0$ by probing the interferometer with photons, and counting the number of times the input is equal to the output. This is associated with the probability  
\begin{align}
    P_= = |p_=|^2,\quad p_= = \langle \vec{n} | Y_m^{(\text{sym})} | \vec{n} \rangle. 
\end{align}
We show in Appendix~\ref{sec:symprob} the probability amplitude is   
\begin{align}
    p_= &= c^n - \binom{n}{2} c^{n-2} s^2 + \sum_{j=2}^{\lfloor n/2 \rfloor} k_{n,2j} c^{n-2j} s^{2j}, 
\end{align}
where the first two terms are always the same irrespective of which modes the photons are injected into, and $k_{n,2j}$ are unimportant coefficients. We will show that these first two terms are the only terms which contribute to the QFI near $\phi\approx0$. Intuitively, this is because $\lim_{\phi\rightarrow 0} s = \lim_{\phi\rightarrow 0} [\sin(\phi/2)/\sqrt{m-1}] = 0$, therefore terms containing higher orders of $s$ are irrelevant when determining $\mathcal{F}^{(\text{sym})}(\phi \approx 0)$. 

We will simplify our Fisher information equation based on the measurement procedure we are considering
\begin{align}
    \text{outcome: input = output} &\Rightarrow P_=(\phi), \\ 
    \text{outcome: input $\neq$ output} &\Rightarrow P_{\neq} (\phi) = 1 - P_=(\phi). 
\end{align}
This means that the derivative of each outcome's probability is related $P_{\neq}' (\phi) = - P_='(\phi)$, so we can simplify
\begin{align}
    \mathcal{F}^{(\text{sym})} \equiv \sum_{o} \frac{(P'_{o})^2}{P_{o}}&= \frac{(P'_=)^2}{P_=} + \frac{(P'_{\neq})^2}{P_{\neq}} = \frac{(P'_=)^2}{P_=(1-P_=)}. \nonumber 
\end{align}
Since the probability amplitude $p_=$ is real, we note that $P_==(p_=)^2$ means $P_='=2p_=p_='$ hence we can get a simple equation for the Fisher information
\begin{align}
    \mathcal{F}^{(\text{sym})} = \frac{4(p_=')^2}{1-p_=^2}.
\end{align}

Now, to get the required expressions for $p_=^2$ and $(p_=')^2$, we will change $s$ to be 
\begin{align}
    t = \sin(\phi/2) = s\sqrt{m-1},
\end{align}
this way we can keep track of the $m$ factors, as well as simplify the derivatives to $t'=c/2$ and $c'=-t/2$. This means we can represent
\begin{align}
    p_= = c^n - \tfrac{\binom{n}{2}}{m-1} c^{n-2} t^2 + \sum_{i=2}^{\lfloor n/2 \rfloor} k_{n,2i}^{(1)} c^{n-2i} t^{2i},
\end{align}
where we will be using these $k$ variables to absorb the coefficients in front of irrelevant terms, where the superscript in brackets is just to label different $k$ values. The squared of the above expression will be
\begin{align}
    p_=^2 = c^{2n} - \tfrac{2\binom{n}{2}}{m-1} c^{2n-2} t^2 + \sum_{i=2}^{2\lfloor n/2 \rfloor} k_{n,2i}^{(2)} c^{2n-2i} t^{2i}.
\end{align}
The derivative can be calculated to be
\begin{align}
    p_=' &= -\left({\scriptstyle \frac{n}{2} + \frac{\binom{n}{2}}{m-1} }\right) c^{n-1} t + \sum_{i=2}^{\lfloor n/2 \rfloor} k_{n,2i}^{(3)} c^{n-2i+1} t^{2i-1}, \nonumber \\ 
    (p_=')^2 &= \left({\scriptstyle \frac{n}{2} + \frac{\binom{n}{2}}{m-1} }\right)^2 c^{2n-2} t^2 + \sum_{i=2}^{2\lfloor n/2 \rfloor-1} k_{n,2i}^{(4)} c^{2n-2i} t^{2i}. \nonumber
\end{align}
Finally, we can calculate the Fisher information as follows
\begin{align}
    \mathcal{F}^{(\text{sym})} &= \tfrac{
    4\left({\scriptstyle \frac{n}{2} + \frac{\binom{n}{2}}{m-1} }\right)^2 c^{2n-2} t^2 + \sum_{i=2}^{2\lfloor n/2 \rfloor-1} k_{n,2i}^{(5)} c^{2n-2i} t^{2i}
    }{
    1-c^{2n} + \tfrac{2\binom{n}{2}}{m-1} c^{2n-2} t^2 + \sum_{i=2}^{2\lfloor n/2 \rfloor} k_{n,2i}^{(6)} c^{2n-2i} t^{2i}
    }\tfrac{1/t^2}{1/t^2}, \nonumber \\
    &= \tfrac{
    4\left({\scriptstyle \frac{n}{2} + \frac{\binom{n}{2}}{m-1} }\right)^2 c^{2n-2} + \sum_{i=2}^{2\lfloor n/2 \rfloor-1} k_{n,2i}^{(5)} c^{2n-2i} t^{2i-2}
    }{
    \tfrac{1-c^{2n}}{t^2} + \tfrac{2\binom{n}{2}}{m-1} c^{2n-2} + \sum_{i=2}^{2\lfloor n/2 \rfloor} k_{n,2i}^{(6)} c^{2n-2i} t^{2i-2}
    }, 
\end{align}
where we added the $1/t^2$ factor because we want a non-zero term in the numerator upon taking the limit. We note that $\lim_{\phi \rightarrow 0} t = 0$, $\lim_{\phi \rightarrow 0} c = 1$ and $\lim_{\phi \rightarrow 0} (1-c^{2n})/t^2 = n$, hence we can calculate the Fisher information as $\phi$ approaches $0$ as follows
\begin{align}
    \lim_{\phi\rightarrow 0}\mathcal{F}^{(\text{sym})}(\phi) = \frac{
    4\left({\scriptstyle \frac{n}{2} + \frac{\binom{n}{2}}{m-1} }\right)^2 
    }{
    n + \tfrac{2\binom{n}{2}}{m-1} 
    }
    = n + \frac{n(n-1)}{m-1}. 
\end{align}
All terms with $k$ coefficients are irrelevant in this limit, hence the Fisher information is the same irrespective of where the $n$ single photons were launched. Furthermore, this metrology apparatus clearly beats the classical shot-noise limit of $n$. This result matches our general QFI calculation in Eq.~\eqref{eq:symF} and hence saturates the precision bound, considering for this single photons scenario that $n_j\in\{0,1\}\Rightarrow \sum^m_{j=1} \sum^m_{k \neq j} n_j n_k = 2\binom{n}{2} = n(n-1)$. 

It is clear that the $Y_m^{(\text{sym})}$ scattering matrix has the appropriate symmetry which equalises the amount of information about the unknown phase amongst all modes, such that it will always give a quantum enhanced detection. However, these results also suggests multimode interferometers with similar phase structure will give roughly the same amount of information after many samples. That said, since the probability of quantum advantage events in the separable interferometer is unlikely, we can expect that the symmetric interferometer will have a statistical advantage at certain finite sample sizes, which we investigate in the next section. 

\section{Contrasting the Interferometers} \label{sec:advan}

\begin{figure*}[htbp]
    \begin{center}
        \includegraphics[width=\linewidth]{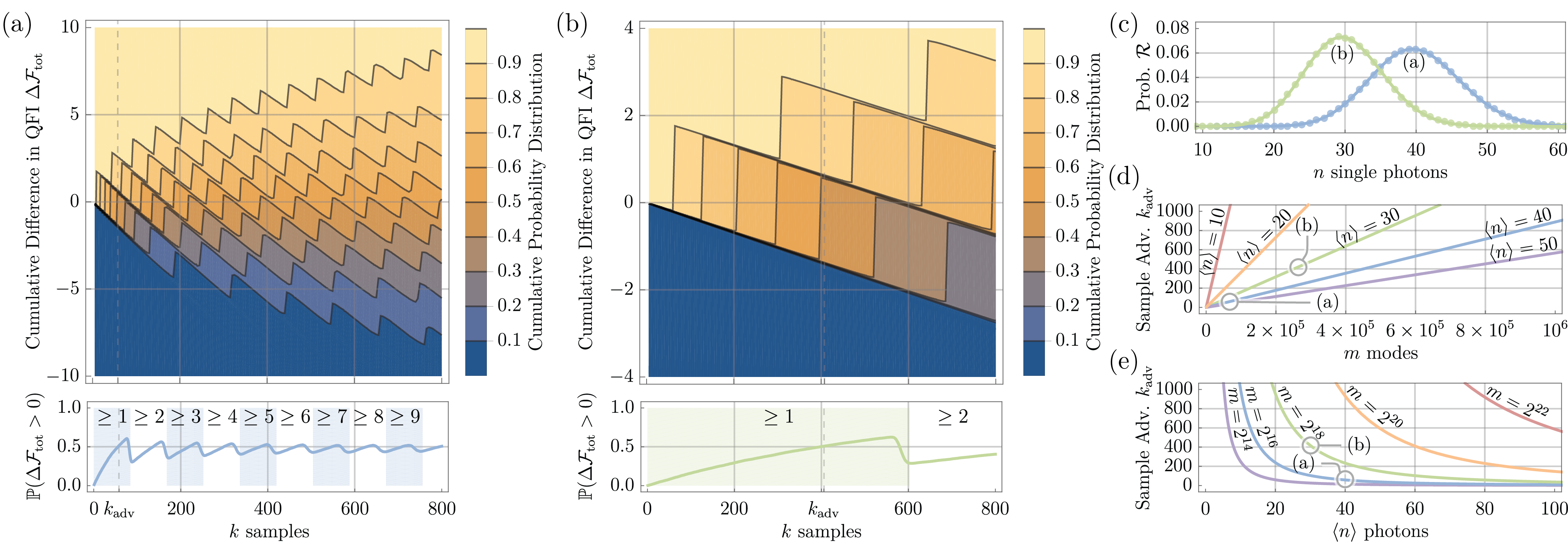}
        \caption{\label{fig:MC} 
            (a) The upper graph is the cumulative probability distribution of $10^4$ random walkers. Each individual walker tracks the total Fisher information difference $\Delta \mathcal{F}_{\text{tot}}(k) = \mathcal{F}^{(\text{sep})}_{\text{tot}}(k) - \mathcal{F}^{(\text{sym})}_{\text{tot}}(k)$, between the separable $Y_m^{(\text{sep})}$ and symmetric interferometers $Y_m^{(\text{sym})}$, as more random samples are generated. The lower graph distills the probability that the separable experiment gave more information $\mathbb{P}[\mathcal{F}^{(\text{sep})}_{\text{tot}}(k) > \mathcal{F}^{(\text{sym})}_{\text{tot}}(k)]$, for a given sample size $k$; the maximum sample size in which $Y_m^{(\text{sym})}$ still has an advantage $k_{\text{adv}}$ is indicated. The shaded regions with labels $\geq1$, $\geq2$, etc, refers to the number of quantum enhanced measurements required by the separable interferometer to beat the symmetric interferometer. These graphs were constructed using a system with $m=2^{16}$ modes and samples from a scattershot source with $\chi\approx 0.0247$ squeezers.
            (b) The same graphs as (a), however with more modes $m=2^{18}$ and weaker squeezing $\chi\approx 0.0107$. 
            (c) The photon statistics of the samples, which shows that these particular $m$ and $\chi$ parameters translates to an average of $\langle n \rangle = 40$ photons for (a) and $\langle n \rangle = 30$ photons for (b). 
            We plot an analytically derived expression for the maximum sample size advantage of the symmetric interferometer $k_{\text{adv}}$, which shows that it (d) increases linearly as we increase the number of modes $m$ in the system, while (e) decreases inversely as we increase the average number of photons $\langle n \rangle$.} 
    \end{center}
\end{figure*} 

Here we aim to quantify the relative performance of the symmetric interferometer in comparison to the separable interferometer, as a function of the sample size. To do this, we implement a Monte Carlo simulation where we randomly generate inputs from an $m$ mode scattershot source, which has the same amount of squeezing $\chi\in[0,1]$ in each mode as shown in Fig.~\ref{fig:scattershot}. The $j$-th mode squeezer creates the following two-mode squeezed state
\begin{align}
    |\psi_j\rangle &= \sqrt{1-\chi^2}\sum_{n_j=0}^{\infty} \chi^{n_j} |n_j n_j \rangle, \\ 
    \mathcal{Q}_{n_j} &= (1-\chi^2) \chi^{2n_j},
\end{align}
where $\mathcal{Q}_{n_j}$ is the probability of generating $n_j$ photons. We use this probability distribution to generate a random sample $|n_1 \cdots n_m\rangle$, and calculate the following 
\begin{align}
    \Delta \mathcal{F} &\equiv \mathcal{F}^{(\text{sep})} - \mathcal{F}^{(\text{sym})}, \nonumber \\ 
    &= 2\sum_{j=1}^{m/2} n_{2j-1} n_{2j} - \frac{1}{m-1} \sum_{j=1}^m \sum_{k \neq j}^m n_j n_k,
\end{align}
which is the difference between the information from the separable and symmetric interferometers. We sequentially generate more random scattershot samples, each time adding to a running total of information difference 
\begin{align}
    \Delta \mathcal{F}_{\text{tot}}(k) \equiv \sum^k_{j=1} \Delta \mathcal{F}_j.
\end{align}
We repeat this entire process multiple times as separate Markov chains or random walkers, which are treated as independent experiments with the same $(m,\chi)$ parameters. From this, we can infer the probability $\mathbb{P}(\Delta \mathcal{F}_{\text{tot}}>0|k)$ in which the separable interferometer will give more information than the symmetric interferometer, for a given $k$ sample size.

The upper graphs of Fig.~\ref{fig:MC}(a) and (b) show the distribution of $10^4$ information difference random walkers, while the lower graphs show the probability that the separable interferometer gave more information. We note Fig.~\ref{fig:MC}(a) was run with a parameter set of $(m = 2^{16}, \chi \approx 0.0247)$ such that the average sample had $\langle n \rangle = 40$ photons, while Fig.~\ref{fig:MC}(b) had a parameter set of $(m = 2^{18}, \chi \approx 0.0107)$ so that $\langle n \rangle = 30$, with the full photon distribution of the samples $\mathcal{R}(n)$ shown in Fig.~\ref{fig:MC}(c). We chose this parameter set in consideration to temporal implementations where creating many modes are inexpensive, since a single source with $m$ time bins can emulate $m$ spatial sources~\cite{motes2014scalable,slussarenko2019photonic}. Furthermore, we only accepted single photon per mode inputs such that the experiment effectively needs just on-off detectors as we only need to identify the outcomes where the input is equal to the output; with low squeezing $\chi$ values, single photons are the most likely outcome in any case. As we will show, this multiple single photons restriction means we can analytically analyse these Monte Carlo results, and ultimately derive an expression for how the symmetric interferometer advantage scales with the experimental parameters.

There is a clear oscillatory behaviour to the graphs of the $\Delta \mathcal{F}_{\text{tot}}(k)$ walker distributions and the $\mathbb{P}(\Delta \mathcal{F}_{\text{tot}}>0|k)$ probabilities. This can be explained because the separable interferometer needs a certain number $x$ of enhanced detection events (i.e. $Y_2|11\rangle$) to beat the symmetric interferometer, for a given range of sample sizes $k$. This is shown in the lower graphs of Fig.~\ref{fig:MC}(a) and (b), where the numbers $\geq 1, \geq 2,$ etc, associated with the shaded regions are explicitly pointing out the minimum $x$ required such that $\Delta \mathcal{F}_{\text{tot}} = \mathcal{F}_{\text{tot}}^{(\text{sep})} - \mathcal{F}_{\text{tot}}^{(\text{sym})} >0$. We analytically calculate these regions as follows 
\begin{align}
    \mathcal{F}_{\text{tot}}^{(\text{sep})} &> \mathcal{F}_{\text{tot}}^{(\text{sym})}, \\ 
    k\langle n \rangle + 2 \sum_{j=1}^k x_j &> k\left(\langle n \rangle + \frac{\langle n \rangle (\langle n \rangle-1)}{m-1}\right), \\ 
    x &> \frac{k \langle n \rangle (\langle n \rangle-1)}{2(m-1)},  
\end{align}
where $x \equiv \sum_{j=1}^k x_j$. This inequality expression is true for the first region $x\geq 1$ when the number of samples is between $k\in\left[1,\frac{2(m-1)}{\langle n \rangle (\langle n \rangle-1)}\right]$, which is $k\in\left[1,84\right]$ and $k\in\left[1,603\right]$ for the parameter set chosen in Fig.~\ref{fig:MC}(a) and (b), respectively. 

The maximum sample size in which the symmetric interferometer still has an advantage $k_{\text{adv}}$, exists in this first $x\geq 1$ region when $\mathbb{P}(\Delta \mathcal{F}_{\text{tot}}>0|k_{\text{adv}}) = 0.5$, and is labelled with dashed vertical lines in the lower graphs. The probability that $x\geq 1$ occurs is equivalent to unity minus the probability that $x = 0$ occurs, which we know from Eq.~\eqref{eq:sepPx} as $\mathcal{P}(x=0) = \binom{m}{n}^{-1} \binom{m/2}{n} 2^{n}$ for one sample. Hence this means that 
\begin{align}
    \mathbb{P}(\Delta \mathcal{F}_{\text{tot}}>0|k) &= 1 - \left(\binom{m}{\langle n \rangle}^{-1} \binom{m/2}{\langle n \rangle} 2^{\langle n \rangle}\right)^k, \nonumber \\  
    0.5 &= 1 - \left(\binom{m}{\langle n \rangle}^{-1} \binom{m/2}{\langle n \rangle} 2^{\langle n \rangle}\right)^{k_{\text{adv}}}, \nonumber \\ 
    \Rightarrow k_{\text{adv}} &= \frac{\log 0.5 }{\log \left(2^{\langle n \rangle} \binom{m/2}{\langle n \rangle} \right)-\log\binom{m}{\langle n \rangle}}. 
\end{align}
We plot this $k_{\text{adv}}$ in Fig.~\ref{fig:MC}(d) and (e), for different $m$ and $\langle n \rangle$ values, which shows that there are certain parameter regimes in which the symmetric interferometer can have a significant advantage over the separable interferometer. While the separable interferometer is easier to implement experimentally, there are situations where the symmetric interferometer is preferable if the number of measurements one can take is limited, such as with photosensitive samples. 

For completion we have also simulated this experiment with a small amount of modes and with inputs that can have more than one photon per mode $n_j\in\mathbb{N}$. This more closely corresponds to an experimental apparatus with spatial modes, and has good squeezing sources with photon number resolving detectors~\cite{arrazola2021quantum}. These results are given in Appendix~\ref{sec:advanbunch}, where we show that the oscillating feature is not as prominent given high enough squeezing, however it has similar scaling features in comparison to the multiple single photons case. 

Finally, we also compare the uniform and symmetric interferometers using a similar Monte Carlo simulation. This is shown in Appendix~\ref{sec:advanuni}, in which we calculated the total information difference between these two interferometers $\Delta \mathcal{F}'_{\text{tot}} = \mathcal{F}_{\text{tot}}^{(\text{uni})} - \mathcal{F}_{\text{tot}}^{(\text{sym})} >0$. It can be seen that the symmetric interferometer also has an advantage over the uniform interferometer, though to a lesser degree in comparison to the the separable interferometer. This confirms what was expected, since the probability of quantum enhanced detection events is the lowest for the separable interferometer, better for the uniform interferometer, and unity for the symmetric interferometer. 

\section{Conclusion} \label{sec:con}

In this article, we investigated how a random photonic (scattershot) source, which has good photon number scaling properties, may be used for quantum metrology purposes. To do this, we introduced three different scalable interferometers with $m$ modes, that all have the same distribution of phase resources. We firstly consider the separable interferometer, which consists of stacking together $m/2$ standard two-mode Mach-Zehnder interferometers. This separable interferometer is a natural baseline to compare other similar interferometers, even at different sizes, as we show that it is metrologically equivalent to just repeat sampling $m/2$ times using a single Mach-Zehnder interferometer. We proved that both the uniform interferometer (made from quantum Fourier transformation devices) and the separable interferometer require the photons to be injected into particular input ports to experience a quantum enhanced detection. In contrast, we show that the symmetric interferometer can always beat the shot-noise limit, irrespective of the input from the scattershot source. However, due to the magnitude of the enhancement, we prove that the amount of information from all three multimode interferometers will be asymptotically (in samples) the same and more than the classical shot-noise limit. To reinforce these results, we detail a specific measurement using single photons that can saturate the same quantum enhanced $\Delta\phi$ precision bound. Finally, we employed Monte Carlo simulations which contrasts all three interferometers for finite sample sizes and in various different experimental regimes; this showed that there are regimes where the symmetric interferometer provides an advantage.

\begin{acknowledgments}
This research was supported by the Australian Research Council Centre of Excellence for Quantum Computation and Communication Technology (Project No. CE170100012).
\end{acknowledgments}

\bibliography{paper}

\begin{thebibliography}{38}%
\makeatletter
\providecommand \@ifxundefined [1]{%
 \@ifx{#1\undefined}
}%
\providecommand \@ifnum [1]{%
 \ifnum #1\expandafter \@firstoftwo
 \else \expandafter \@secondoftwo
 \fi
}%
\providecommand \@ifx [1]{%
 \ifx #1\expandafter \@firstoftwo
 \else \expandafter \@secondoftwo
 \fi
}%
\providecommand \natexlab [1]{#1}%
\providecommand \enquote  [1]{``#1''}%
\providecommand \bibnamefont  [1]{#1}%
\providecommand \bibfnamefont [1]{#1}%
\providecommand \citenamefont [1]{#1}%
\providecommand \href@noop [0]{\@secondoftwo}%
\providecommand \href [0]{\begingroup \@sanitize@url \@href}%
\providecommand \@href[1]{\@@startlink{#1}\@@href}%
\providecommand \@@href[1]{\endgroup#1\@@endlink}%
\providecommand \@sanitize@url [0]{\catcode `\\12\catcode `\$12\catcode
  `\&12\catcode `\#12\catcode `\^12\catcode `\_12\catcode `\%12\relax}%
\providecommand \@@startlink[1]{}%
\providecommand \@@endlink[0]{}%
\providecommand \url  [0]{\begingroup\@sanitize@url \@url }%
\providecommand \@url [1]{\endgroup\@href {#1}{\urlprefix }}%
\providecommand \urlprefix  [0]{URL }%
\providecommand \Eprint [0]{\href }%
\providecommand \doibase [0]{http://dx.doi.org/}%
\providecommand \selectlanguage [0]{\@gobble}%
\providecommand \bibinfo  [0]{\@secondoftwo}%
\providecommand \bibfield  [0]{\@secondoftwo}%
\providecommand \translation [1]{[#1]}%
\providecommand \BibitemOpen [0]{}%
\providecommand \bibitemStop [0]{}%
\providecommand \bibitemNoStop [0]{.\EOS\space}%
\providecommand \EOS [0]{\spacefactor3000\relax}%
\providecommand \BibitemShut  [1]{\csname bibitem#1\endcsname}%
\let\auto@bib@innerbib\@empty
%</preamble>
\bibitem [{\citenamefont {Giovannetti}\ \emph {et~al.}(2006)\citenamefont
  {Giovannetti}, \citenamefont {Lloyd},\ and\ \citenamefont
  {Maccone}}]{giovannetti2006quantum}%
  \BibitemOpen
  \bibfield  {author} {\bibinfo {author} {\bibfnamefont {V.}~\bibnamefont
  {Giovannetti}}, \bibinfo {author} {\bibfnamefont {S.}~\bibnamefont {Lloyd}},
  \ and\ \bibinfo {author} {\bibfnamefont {L.}~\bibnamefont {Maccone}},\
  }\href@noop {} {\bibfield  {journal} {\bibinfo  {journal} {Physical review
  letters}\ }\textbf {\bibinfo {volume} {96}},\ \bibinfo {pages} {010401}
  (\bibinfo {year} {2006})}\BibitemShut {NoStop}%
\bibitem [{\citenamefont {Giovannetti}\ \emph {et~al.}(2011)\citenamefont
  {Giovannetti}, \citenamefont {Lloyd},\ and\ \citenamefont
  {Maccone}}]{giovannetti2011advances}%
  \BibitemOpen
  \bibfield  {author} {\bibinfo {author} {\bibfnamefont {V.}~\bibnamefont
  {Giovannetti}}, \bibinfo {author} {\bibfnamefont {S.}~\bibnamefont {Lloyd}},
  \ and\ \bibinfo {author} {\bibfnamefont {L.}~\bibnamefont {Maccone}},\
  }\href@noop {} {\bibfield  {journal} {\bibinfo  {journal} {Nature photonics}\
  }\textbf {\bibinfo {volume} {5}},\ \bibinfo {pages} {222} (\bibinfo {year}
  {2011})}\BibitemShut {NoStop}%
\bibitem [{\citenamefont {Bollinger}\ \emph {et~al.}(1996)\citenamefont
  {Bollinger}, \citenamefont {Itano}, \citenamefont {Wineland},\ and\
  \citenamefont {Heinzen}}]{bollinger1996optimal}%
  \BibitemOpen
  \bibfield  {author} {\bibinfo {author} {\bibfnamefont {J.~J.}\ \bibnamefont
  {Bollinger}}, \bibinfo {author} {\bibfnamefont {W.~M.}\ \bibnamefont
  {Itano}}, \bibinfo {author} {\bibfnamefont {D.~J.}\ \bibnamefont {Wineland}},
  \ and\ \bibinfo {author} {\bibfnamefont {D.~J.}\ \bibnamefont {Heinzen}},\
  }\href@noop {} {\bibfield  {journal} {\bibinfo  {journal} {Physical Review
  A}\ }\textbf {\bibinfo {volume} {54}},\ \bibinfo {pages} {R4649} (\bibinfo
  {year} {1996})}\BibitemShut {NoStop}%
\bibitem [{\citenamefont {Lee}\ \emph {et~al.}(2002)\citenamefont {Lee},
  \citenamefont {Kok},\ and\ \citenamefont {Dowling}}]{lee2002quantum}%
  \BibitemOpen
  \bibfield  {author} {\bibinfo {author} {\bibfnamefont {H.}~\bibnamefont
  {Lee}}, \bibinfo {author} {\bibfnamefont {P.}~\bibnamefont {Kok}}, \ and\
  \bibinfo {author} {\bibfnamefont {J.~P.}\ \bibnamefont {Dowling}},\
  }\href@noop {} {\bibfield  {journal} {\bibinfo  {journal} {Journal of Modern
  Optics}\ }\textbf {\bibinfo {volume} {49}},\ \bibinfo {pages} {2325}
  (\bibinfo {year} {2002})}\BibitemShut {NoStop}%
\bibitem [{\citenamefont {Joo}\ \emph {et~al.}(2011)\citenamefont {Joo},
  \citenamefont {Munro},\ and\ \citenamefont {Spiller}}]{joo2011quantum}%
  \BibitemOpen
  \bibfield  {author} {\bibinfo {author} {\bibfnamefont {J.}~\bibnamefont
  {Joo}}, \bibinfo {author} {\bibfnamefont {W.~J.}\ \bibnamefont {Munro}}, \
  and\ \bibinfo {author} {\bibfnamefont {T.~P.}\ \bibnamefont {Spiller}},\
  }\href@noop {} {\bibfield  {journal} {\bibinfo  {journal} {Physical review
  letters}\ }\textbf {\bibinfo {volume} {107}},\ \bibinfo {pages} {083601}
  (\bibinfo {year} {2011})}\BibitemShut {NoStop}%
\bibitem [{\citenamefont {Tillmann}\ \emph {et~al.}(2013)\citenamefont
  {Tillmann}, \citenamefont {Daki{\'c}}, \citenamefont {Heilmann},
  \citenamefont {Nolte}, \citenamefont {Szameit},\ and\ \citenamefont
  {Walther}}]{tillmann2013experimental}%
  \BibitemOpen
  \bibfield  {author} {\bibinfo {author} {\bibfnamefont {M.}~\bibnamefont
  {Tillmann}}, \bibinfo {author} {\bibfnamefont {B.}~\bibnamefont {Daki{\'c}}},
  \bibinfo {author} {\bibfnamefont {R.}~\bibnamefont {Heilmann}}, \bibinfo
  {author} {\bibfnamefont {S.}~\bibnamefont {Nolte}}, \bibinfo {author}
  {\bibfnamefont {A.}~\bibnamefont {Szameit}}, \ and\ \bibinfo {author}
  {\bibfnamefont {P.}~\bibnamefont {Walther}},\ }\href@noop {} {\bibfield
  {journal} {\bibinfo  {journal} {Nature photonics}\ }\textbf {\bibinfo
  {volume} {7}},\ \bibinfo {pages} {540} (\bibinfo {year} {2013})}\BibitemShut
  {NoStop}%
\bibitem [{\citenamefont {Spring}\ \emph {et~al.}(2013)\citenamefont {Spring},
  \citenamefont {Metcalf}, \citenamefont {Humphreys}, \citenamefont
  {Kolthammer}, \citenamefont {Jin}, \citenamefont {Barbieri}, \citenamefont
  {Datta}, \citenamefont {Thomas-Peter}, \citenamefont {Langford},
  \citenamefont {Kundys} \emph {et~al.}}]{spring2013boson}%
  \BibitemOpen
  \bibfield  {author} {\bibinfo {author} {\bibfnamefont {J.~B.}\ \bibnamefont
  {Spring}}, \bibinfo {author} {\bibfnamefont {B.~J.}\ \bibnamefont {Metcalf}},
  \bibinfo {author} {\bibfnamefont {P.~C.}\ \bibnamefont {Humphreys}}, \bibinfo
  {author} {\bibfnamefont {W.~S.}\ \bibnamefont {Kolthammer}}, \bibinfo
  {author} {\bibfnamefont {X.-M.}\ \bibnamefont {Jin}}, \bibinfo {author}
  {\bibfnamefont {M.}~\bibnamefont {Barbieri}}, \bibinfo {author}
  {\bibfnamefont {A.}~\bibnamefont {Datta}}, \bibinfo {author} {\bibfnamefont
  {N.}~\bibnamefont {Thomas-Peter}}, \bibinfo {author} {\bibfnamefont {N.~K.}\
  \bibnamefont {Langford}}, \bibinfo {author} {\bibfnamefont {D.}~\bibnamefont
  {Kundys}},  \emph {et~al.},\ }\href@noop {} {\bibfield  {journal} {\bibinfo
  {journal} {Science}\ }\textbf {\bibinfo {volume} {339}},\ \bibinfo {pages}
  {798} (\bibinfo {year} {2013})}\BibitemShut {NoStop}%
\bibitem [{\citenamefont {Broome}\ \emph {et~al.}(2013)\citenamefont {Broome},
  \citenamefont {Fedrizzi}, \citenamefont {Rahimi-Keshari}, \citenamefont
  {Dove}, \citenamefont {Aaronson}, \citenamefont {Ralph},\ and\ \citenamefont
  {White}}]{broome2013photonic}%
  \BibitemOpen
  \bibfield  {author} {\bibinfo {author} {\bibfnamefont {M.~A.}\ \bibnamefont
  {Broome}}, \bibinfo {author} {\bibfnamefont {A.}~\bibnamefont {Fedrizzi}},
  \bibinfo {author} {\bibfnamefont {S.}~\bibnamefont {Rahimi-Keshari}},
  \bibinfo {author} {\bibfnamefont {J.}~\bibnamefont {Dove}}, \bibinfo {author}
  {\bibfnamefont {S.}~\bibnamefont {Aaronson}}, \bibinfo {author}
  {\bibfnamefont {T.~C.}\ \bibnamefont {Ralph}}, \ and\ \bibinfo {author}
  {\bibfnamefont {A.~G.}\ \bibnamefont {White}},\ }\href@noop {} {\bibfield
  {journal} {\bibinfo  {journal} {Science}\ }\textbf {\bibinfo {volume}
  {339}},\ \bibinfo {pages} {794} (\bibinfo {year} {2013})}\BibitemShut
  {NoStop}%
\bibitem [{\citenamefont {Aaronson}\ and\ \citenamefont
  {Arkhipov}(2011)}]{aaronson2011computational}%
  \BibitemOpen
  \bibfield  {author} {\bibinfo {author} {\bibfnamefont {S.}~\bibnamefont
  {Aaronson}}\ and\ \bibinfo {author} {\bibfnamefont {A.}~\bibnamefont
  {Arkhipov}},\ }in\ \href@noop {} {\emph {\bibinfo {booktitle} {Proceedings of
  the forty-third annual ACM symposium on Theory of computing}}}\ (\bibinfo
  {year} {2011})\ pp.\ \bibinfo {pages} {333--342}\BibitemShut {NoStop}%
\bibitem [{\citenamefont {Zhong}\ \emph {et~al.}(2020)\citenamefont {Zhong},
  \citenamefont {Wang}, \citenamefont {Deng}, \citenamefont {Chen},
  \citenamefont {Peng}, \citenamefont {Luo}, \citenamefont {Qin}, \citenamefont
  {Wu}, \citenamefont {Ding}, \citenamefont {Hu} \emph
  {et~al.}}]{zhong2020quantum}%
  \BibitemOpen
  \bibfield  {author} {\bibinfo {author} {\bibfnamefont {H.-S.}\ \bibnamefont
  {Zhong}}, \bibinfo {author} {\bibfnamefont {H.}~\bibnamefont {Wang}},
  \bibinfo {author} {\bibfnamefont {Y.-H.}\ \bibnamefont {Deng}}, \bibinfo
  {author} {\bibfnamefont {M.-C.}\ \bibnamefont {Chen}}, \bibinfo {author}
  {\bibfnamefont {L.-C.}\ \bibnamefont {Peng}}, \bibinfo {author}
  {\bibfnamefont {Y.-H.}\ \bibnamefont {Luo}}, \bibinfo {author} {\bibfnamefont
  {J.}~\bibnamefont {Qin}}, \bibinfo {author} {\bibfnamefont {D.}~\bibnamefont
  {Wu}}, \bibinfo {author} {\bibfnamefont {X.}~\bibnamefont {Ding}}, \bibinfo
  {author} {\bibfnamefont {Y.}~\bibnamefont {Hu}},  \emph {et~al.},\
  }\href@noop {} {\bibfield  {journal} {\bibinfo  {journal} {Science}\ }\textbf
  {\bibinfo {volume} {370}},\ \bibinfo {pages} {1460} (\bibinfo {year}
  {2020})}\BibitemShut {NoStop}%
\bibitem [{\citenamefont {Motes}\ \emph {et~al.}(2015)\citenamefont {Motes},
  \citenamefont {Olson}, \citenamefont {Rabeaux}, \citenamefont {Dowling},
  \citenamefont {Olson},\ and\ \citenamefont {Rohde}}]{motes2015linear}%
  \BibitemOpen
  \bibfield  {author} {\bibinfo {author} {\bibfnamefont {K.~R.}\ \bibnamefont
  {Motes}}, \bibinfo {author} {\bibfnamefont {J.~P.}\ \bibnamefont {Olson}},
  \bibinfo {author} {\bibfnamefont {E.~J.}\ \bibnamefont {Rabeaux}}, \bibinfo
  {author} {\bibfnamefont {J.~P.}\ \bibnamefont {Dowling}}, \bibinfo {author}
  {\bibfnamefont {S.~J.}\ \bibnamefont {Olson}}, \ and\ \bibinfo {author}
  {\bibfnamefont {P.~P.}\ \bibnamefont {Rohde}},\ }\href@noop {} {\bibfield
  {journal} {\bibinfo  {journal} {Physical review letters}\ }\textbf {\bibinfo
  {volume} {114}},\ \bibinfo {pages} {170802} (\bibinfo {year}
  {2015})}\BibitemShut {NoStop}%
\bibitem [{\citenamefont {Olson}\ \emph {et~al.}(2017)\citenamefont {Olson},
  \citenamefont {Motes}, \citenamefont {Birchall}, \citenamefont {Studer},
  \citenamefont {LaBorde}, \citenamefont {Moulder}, \citenamefont {Rohde},\
  and\ \citenamefont {Dowling}}]{olson2017linear}%
  \BibitemOpen
  \bibfield  {author} {\bibinfo {author} {\bibfnamefont {J.~P.}\ \bibnamefont
  {Olson}}, \bibinfo {author} {\bibfnamefont {K.~R.}\ \bibnamefont {Motes}},
  \bibinfo {author} {\bibfnamefont {P.~M.}\ \bibnamefont {Birchall}}, \bibinfo
  {author} {\bibfnamefont {N.~M.}\ \bibnamefont {Studer}}, \bibinfo {author}
  {\bibfnamefont {M.}~\bibnamefont {LaBorde}}, \bibinfo {author} {\bibfnamefont
  {T.}~\bibnamefont {Moulder}}, \bibinfo {author} {\bibfnamefont {P.~P.}\
  \bibnamefont {Rohde}}, \ and\ \bibinfo {author} {\bibfnamefont {J.~P.}\
  \bibnamefont {Dowling}},\ }\href@noop {} {\bibfield  {journal} {\bibinfo
  {journal} {Physical Review A}\ }\textbf {\bibinfo {volume} {96}},\ \bibinfo
  {pages} {013810} (\bibinfo {year} {2017})}\BibitemShut {NoStop}%
\bibitem [{\citenamefont {Su}\ \emph {et~al.}(2017)\citenamefont {Su},
  \citenamefont {Li}, \citenamefont {Rohde}, \citenamefont {Huang},
  \citenamefont {Wang}, \citenamefont {Li}, \citenamefont {Liu}, \citenamefont
  {Dowling}, \citenamefont {Lu},\ and\ \citenamefont
  {Pan}}]{su2017multiphoton}%
  \BibitemOpen
  \bibfield  {author} {\bibinfo {author} {\bibfnamefont {Z.-E.}\ \bibnamefont
  {Su}}, \bibinfo {author} {\bibfnamefont {Y.}~\bibnamefont {Li}}, \bibinfo
  {author} {\bibfnamefont {P.~P.}\ \bibnamefont {Rohde}}, \bibinfo {author}
  {\bibfnamefont {H.-L.}\ \bibnamefont {Huang}}, \bibinfo {author}
  {\bibfnamefont {X.-L.}\ \bibnamefont {Wang}}, \bibinfo {author}
  {\bibfnamefont {L.}~\bibnamefont {Li}}, \bibinfo {author} {\bibfnamefont
  {N.-L.}\ \bibnamefont {Liu}}, \bibinfo {author} {\bibfnamefont {J.~P.}\
  \bibnamefont {Dowling}}, \bibinfo {author} {\bibfnamefont {C.-Y.}\
  \bibnamefont {Lu}}, \ and\ \bibinfo {author} {\bibfnamefont {J.-W.}\
  \bibnamefont {Pan}},\ }\href@noop {} {\bibfield  {journal} {\bibinfo
  {journal} {Physical review letters}\ }\textbf {\bibinfo {volume} {119}},\
  \bibinfo {pages} {080502} (\bibinfo {year} {2017})}\BibitemShut {NoStop}%
\bibitem [{\citenamefont {Lund}\ \emph {et~al.}(2014)\citenamefont {Lund},
  \citenamefont {Laing}, \citenamefont {Rahimi-Keshari}, \citenamefont
  {Rudolph}, \citenamefont {O’Brien},\ and\ \citenamefont
  {Ralph}}]{lund2014boson}%
  \BibitemOpen
  \bibfield  {author} {\bibinfo {author} {\bibfnamefont {A.~P.}\ \bibnamefont
  {Lund}}, \bibinfo {author} {\bibfnamefont {A.}~\bibnamefont {Laing}},
  \bibinfo {author} {\bibfnamefont {S.}~\bibnamefont {Rahimi-Keshari}},
  \bibinfo {author} {\bibfnamefont {T.}~\bibnamefont {Rudolph}}, \bibinfo
  {author} {\bibfnamefont {J.~L.}\ \bibnamefont {O’Brien}}, \ and\ \bibinfo
  {author} {\bibfnamefont {T.~C.}\ \bibnamefont {Ralph}},\ }\href@noop {}
  {\bibfield  {journal} {\bibinfo  {journal} {Physical review letters}\
  }\textbf {\bibinfo {volume} {113}},\ \bibinfo {pages} {100502} (\bibinfo
  {year} {2014})}\BibitemShut {NoStop}%
\bibitem [{\citenamefont {Bentivegna}\ \emph {et~al.}(2015)\citenamefont
  {Bentivegna}, \citenamefont {Spagnolo}, \citenamefont {Vitelli},
  \citenamefont {Flamini}, \citenamefont {Viggianiello}, \citenamefont
  {Latmiral}, \citenamefont {Mataloni}, \citenamefont {Brod}, \citenamefont
  {Galv{\~a}o}, \citenamefont {Crespi} \emph
  {et~al.}}]{bentivegna2015experimental}%
  \BibitemOpen
  \bibfield  {author} {\bibinfo {author} {\bibfnamefont {M.}~\bibnamefont
  {Bentivegna}}, \bibinfo {author} {\bibfnamefont {N.}~\bibnamefont
  {Spagnolo}}, \bibinfo {author} {\bibfnamefont {C.}~\bibnamefont {Vitelli}},
  \bibinfo {author} {\bibfnamefont {F.}~\bibnamefont {Flamini}}, \bibinfo
  {author} {\bibfnamefont {N.}~\bibnamefont {Viggianiello}}, \bibinfo {author}
  {\bibfnamefont {L.}~\bibnamefont {Latmiral}}, \bibinfo {author}
  {\bibfnamefont {P.}~\bibnamefont {Mataloni}}, \bibinfo {author}
  {\bibfnamefont {D.~J.}\ \bibnamefont {Brod}}, \bibinfo {author}
  {\bibfnamefont {E.~F.}\ \bibnamefont {Galv{\~a}o}}, \bibinfo {author}
  {\bibfnamefont {A.}~\bibnamefont {Crespi}},  \emph {et~al.},\ }\href@noop {}
  {\bibfield  {journal} {\bibinfo  {journal} {Science advances}\ }\textbf
  {\bibinfo {volume} {1}},\ \bibinfo {pages} {e1400255} (\bibinfo {year}
  {2015})}\BibitemShut {NoStop}%
\bibitem [{\citenamefont {Zhong}\ \emph {et~al.}(2018)\citenamefont {Zhong},
  \citenamefont {Li}, \citenamefont {Li}, \citenamefont {Peng}, \citenamefont
  {Su}, \citenamefont {Hu}, \citenamefont {He}, \citenamefont {Ding},
  \citenamefont {Zhang}, \citenamefont {Li} \emph {et~al.}}]{zhong201812}%
  \BibitemOpen
  \bibfield  {author} {\bibinfo {author} {\bibfnamefont {H.-S.}\ \bibnamefont
  {Zhong}}, \bibinfo {author} {\bibfnamefont {Y.}~\bibnamefont {Li}}, \bibinfo
  {author} {\bibfnamefont {W.}~\bibnamefont {Li}}, \bibinfo {author}
  {\bibfnamefont {L.-C.}\ \bibnamefont {Peng}}, \bibinfo {author}
  {\bibfnamefont {Z.-E.}\ \bibnamefont {Su}}, \bibinfo {author} {\bibfnamefont
  {Y.}~\bibnamefont {Hu}}, \bibinfo {author} {\bibfnamefont {Y.-M.}\
  \bibnamefont {He}}, \bibinfo {author} {\bibfnamefont {X.}~\bibnamefont
  {Ding}}, \bibinfo {author} {\bibfnamefont {W.}~\bibnamefont {Zhang}},
  \bibinfo {author} {\bibfnamefont {H.}~\bibnamefont {Li}},  \emph {et~al.},\
  }\href@noop {} {\bibfield  {journal} {\bibinfo  {journal} {Physical review
  letters}\ }\textbf {\bibinfo {volume} {121}},\ \bibinfo {pages} {250505}
  (\bibinfo {year} {2018})}\BibitemShut {NoStop}%
\bibitem [{\citenamefont {You}\ \emph {et~al.}(2017)\citenamefont {You},
  \citenamefont {Adhikari}, \citenamefont {Chi}, \citenamefont {LaBorde},
  \citenamefont {Matyas}, \citenamefont {Zhang}, \citenamefont {Su},
  \citenamefont {Byrnes}, \citenamefont {Lu}, \citenamefont {Dowling} \emph
  {et~al.}}]{you2017multiparameter}%
  \BibitemOpen
  \bibfield  {author} {\bibinfo {author} {\bibfnamefont {C.}~\bibnamefont
  {You}}, \bibinfo {author} {\bibfnamefont {S.}~\bibnamefont {Adhikari}},
  \bibinfo {author} {\bibfnamefont {Y.}~\bibnamefont {Chi}}, \bibinfo {author}
  {\bibfnamefont {M.~L.}\ \bibnamefont {LaBorde}}, \bibinfo {author}
  {\bibfnamefont {C.~T.}\ \bibnamefont {Matyas}}, \bibinfo {author}
  {\bibfnamefont {C.}~\bibnamefont {Zhang}}, \bibinfo {author} {\bibfnamefont
  {Z.}~\bibnamefont {Su}}, \bibinfo {author} {\bibfnamefont {T.}~\bibnamefont
  {Byrnes}}, \bibinfo {author} {\bibfnamefont {C.}~\bibnamefont {Lu}}, \bibinfo
  {author} {\bibfnamefont {J.~P.}\ \bibnamefont {Dowling}},  \emph {et~al.},\
  }\href@noop {} {\bibfield  {journal} {\bibinfo  {journal} {Journal of
  Optics}\ }\textbf {\bibinfo {volume} {19}},\ \bibinfo {pages} {124002}
  (\bibinfo {year} {2017})}\BibitemShut {NoStop}%
\bibitem [{\citenamefont {Reck}\ \emph {et~al.}(1994)\citenamefont {Reck},
  \citenamefont {Zeilinger}, \citenamefont {Bernstein},\ and\ \citenamefont
  {Bertani}}]{reck1994experimental}%
  \BibitemOpen
  \bibfield  {author} {\bibinfo {author} {\bibfnamefont {M.}~\bibnamefont
  {Reck}}, \bibinfo {author} {\bibfnamefont {A.}~\bibnamefont {Zeilinger}},
  \bibinfo {author} {\bibfnamefont {H.~J.}\ \bibnamefont {Bernstein}}, \ and\
  \bibinfo {author} {\bibfnamefont {P.}~\bibnamefont {Bertani}},\ }\href@noop
  {} {\bibfield  {journal} {\bibinfo  {journal} {Physical Review Letters}\
  }\textbf {\bibinfo {volume} {73}},\ \bibinfo {pages} {58} (\bibinfo {year}
  {1994})}\BibitemShut {NoStop}%
\bibitem [{\citenamefont {Nielsen}\ and\ \citenamefont
  {Chuang}(2010)}]{nielsen2010quantum}%
  \BibitemOpen
  \bibfield  {author} {\bibinfo {author} {\bibfnamefont {M.~A.}\ \bibnamefont
  {Nielsen}}\ and\ \bibinfo {author} {\bibfnamefont {I.}~\bibnamefont
  {Chuang}},\ }\href@noop {} {\emph {\bibinfo {title} {Quantum Computation and
  Quantum Information}}}\ (\bibinfo  {publisher} {Cambridge University Press},\
  \bibinfo {year} {2010})\BibitemShut {NoStop}%
\bibitem [{\citenamefont {Clements}\ \emph {et~al.}(2016)\citenamefont
  {Clements}, \citenamefont {Humphreys}, \citenamefont {Metcalf}, \citenamefont
  {Kolthammer},\ and\ \citenamefont {Walmsley}}]{clements2016optimal}%
  \BibitemOpen
  \bibfield  {author} {\bibinfo {author} {\bibfnamefont {W.~R.}\ \bibnamefont
  {Clements}}, \bibinfo {author} {\bibfnamefont {P.~C.}\ \bibnamefont
  {Humphreys}}, \bibinfo {author} {\bibfnamefont {B.~J.}\ \bibnamefont
  {Metcalf}}, \bibinfo {author} {\bibfnamefont {W.~S.}\ \bibnamefont
  {Kolthammer}}, \ and\ \bibinfo {author} {\bibfnamefont {I.~A.}\ \bibnamefont
  {Walmsley}},\ }\href@noop {} {\bibfield  {journal} {\bibinfo  {journal}
  {Optica}\ }\textbf {\bibinfo {volume} {3}},\ \bibinfo {pages} {1460}
  (\bibinfo {year} {2016})}\BibitemShut {NoStop}%
\bibitem [{Note1()}]{Note1}%
  \BibitemOpen
  \bibinfo {note} {Whilst the phase parameter $\theta $ is variable, it is not
  considered to be changed whilst photons are evolving through the network.
  Rather, it is considered to be set and fixed before samples are recorded.
  Therefore the presence of this controllable phase shift is not in
  contradiction to our ``passive network'' premise.}\BibitemShut {Stop}%
\bibitem [{\citenamefont {Takeoka}\ \emph {et~al.}(2017)\citenamefont
  {Takeoka}, \citenamefont {Seshadreesan}, \citenamefont {You}, \citenamefont
  {Izumi},\ and\ \citenamefont {Dowling}}]{takeoka2017fundamental}%
  \BibitemOpen
  \bibfield  {author} {\bibinfo {author} {\bibfnamefont {M.}~\bibnamefont
  {Takeoka}}, \bibinfo {author} {\bibfnamefont {K.~P.}\ \bibnamefont
  {Seshadreesan}}, \bibinfo {author} {\bibfnamefont {C.}~\bibnamefont {You}},
  \bibinfo {author} {\bibfnamefont {S.}~\bibnamefont {Izumi}}, \ and\ \bibinfo
  {author} {\bibfnamefont {J.~P.}\ \bibnamefont {Dowling}},\ }\href@noop {}
  {\bibfield  {journal} {\bibinfo  {journal} {Physical Review A}\ }\textbf
  {\bibinfo {volume} {96}},\ \bibinfo {pages} {052118} (\bibinfo {year}
  {2017})}\BibitemShut {NoStop}%
\bibitem [{\citenamefont {Helstrom}\ and\ \citenamefont
  {Helstrom}(1976)}]{helstrom1976quantum}%
  \BibitemOpen
  \bibfield  {author} {\bibinfo {author} {\bibfnamefont {C.~W.}\ \bibnamefont
  {Helstrom}}\ and\ \bibinfo {author} {\bibfnamefont {C.~W.}\ \bibnamefont
  {Helstrom}},\ }\href@noop {} {\emph {\bibinfo {title} {Quantum detection and
  estimation theory}}},\ Vol.~\bibinfo {volume} {84}\ (\bibinfo  {publisher}
  {Academic press New York},\ \bibinfo {year} {1976})\BibitemShut {NoStop}%
\bibitem [{\citenamefont {Escher}\ \emph {et~al.}(2011)\citenamefont {Escher},
  \citenamefont {de~Matos~Filho},\ and\ \citenamefont
  {Davidovich}}]{escher2011general}%
  \BibitemOpen
  \bibfield  {author} {\bibinfo {author} {\bibfnamefont {B.}~\bibnamefont
  {Escher}}, \bibinfo {author} {\bibfnamefont {R.}~\bibnamefont
  {de~Matos~Filho}}, \ and\ \bibinfo {author} {\bibfnamefont {L.}~\bibnamefont
  {Davidovich}},\ }\href@noop {} {\bibfield  {journal} {\bibinfo  {journal}
  {Nature Physics}\ }\textbf {\bibinfo {volume} {7}},\ \bibinfo {pages} {406}
  (\bibinfo {year} {2011})}\BibitemShut {NoStop}%
\bibitem [{\citenamefont {Demkowicz-Dobrza{\'n}ski}\ \emph
  {et~al.}(2012)\citenamefont {Demkowicz-Dobrza{\'n}ski}, \citenamefont
  {Ko{\l}ody{\'n}ski},\ and\ \citenamefont
  {Gu{\c{t}}{\u{a}}}}]{demkowicz2012elusive}%
  \BibitemOpen
  \bibfield  {author} {\bibinfo {author} {\bibfnamefont {R.}~\bibnamefont
  {Demkowicz-Dobrza{\'n}ski}}, \bibinfo {author} {\bibfnamefont
  {J.}~\bibnamefont {Ko{\l}ody{\'n}ski}}, \ and\ \bibinfo {author}
  {\bibfnamefont {M.}~\bibnamefont {Gu{\c{t}}{\u{a}}}},\ }\href@noop {}
  {\bibfield  {journal} {\bibinfo  {journal} {Nature communications}\ }\textbf
  {\bibinfo {volume} {3}},\ \bibinfo {pages} {1} (\bibinfo {year}
  {2012})}\BibitemShut {NoStop}%
\bibitem [{\citenamefont {Braunstein}\ and\ \citenamefont
  {Caves}(1994)}]{braunstein1994statistical}%
  \BibitemOpen
  \bibfield  {author} {\bibinfo {author} {\bibfnamefont {S.~L.}\ \bibnamefont
  {Braunstein}}\ and\ \bibinfo {author} {\bibfnamefont {C.~M.}\ \bibnamefont
  {Caves}},\ }\href@noop {} {\bibfield  {journal} {\bibinfo  {journal}
  {Physical Review Letters}\ }\textbf {\bibinfo {volume} {72}},\ \bibinfo
  {pages} {3439} (\bibinfo {year} {1994})}\BibitemShut {NoStop}%
\bibitem [{\citenamefont {Perarnau-Llobet}\ \emph {et~al.}(2020)\citenamefont
  {Perarnau-Llobet}, \citenamefont {Gonz{\'a}lez-Tudela},\ and\ \citenamefont
  {Cirac}}]{perarnau2020multimode}%
  \BibitemOpen
  \bibfield  {author} {\bibinfo {author} {\bibfnamefont {M.}~\bibnamefont
  {Perarnau-Llobet}}, \bibinfo {author} {\bibfnamefont {A.}~\bibnamefont
  {Gonz{\'a}lez-Tudela}}, \ and\ \bibinfo {author} {\bibfnamefont {J.~I.}\
  \bibnamefont {Cirac}},\ }\href@noop {} {\bibfield  {journal} {\bibinfo
  {journal} {Quantum Science and Technology}\ }\textbf {\bibinfo {volume}
  {5}},\ \bibinfo {pages} {025003} (\bibinfo {year} {2020})}\BibitemShut
  {NoStop}%
\bibitem [{\citenamefont {Lang}\ and\ \citenamefont
  {Caves}(2014)}]{lang2014optimal}%
  \BibitemOpen
  \bibfield  {author} {\bibinfo {author} {\bibfnamefont {M.~D.}\ \bibnamefont
  {Lang}}\ and\ \bibinfo {author} {\bibfnamefont {C.~M.}\ \bibnamefont
  {Caves}},\ }\href@noop {} {\bibfield  {journal} {\bibinfo  {journal}
  {Physical Review A}\ }\textbf {\bibinfo {volume} {90}},\ \bibinfo {pages}
  {025802} (\bibinfo {year} {2014})}\BibitemShut {NoStop}%
\bibitem [{\citenamefont {Yu}\ \emph {et~al.}(2018)\citenamefont {Yu},
  \citenamefont {Zhao}, \citenamefont {Shen}, \citenamefont {Shao},
  \citenamefont {Liu},\ and\ \citenamefont {Wang}}]{yu2018maximal}%
  \BibitemOpen
  \bibfield  {author} {\bibinfo {author} {\bibfnamefont {X.}~\bibnamefont
  {Yu}}, \bibinfo {author} {\bibfnamefont {X.}~\bibnamefont {Zhao}}, \bibinfo
  {author} {\bibfnamefont {L.}~\bibnamefont {Shen}}, \bibinfo {author}
  {\bibfnamefont {Y.}~\bibnamefont {Shao}}, \bibinfo {author} {\bibfnamefont
  {J.}~\bibnamefont {Liu}}, \ and\ \bibinfo {author} {\bibfnamefont
  {X.}~\bibnamefont {Wang}},\ }\href@noop {} {\bibfield  {journal} {\bibinfo
  {journal} {Optics express}\ }\textbf {\bibinfo {volume} {26}},\ \bibinfo
  {pages} {16292} (\bibinfo {year} {2018})}\BibitemShut {NoStop}%
\bibitem [{\citenamefont {Holland}\ and\ \citenamefont
  {Burnett}(1993)}]{holland1993interferometric}%
  \BibitemOpen
  \bibfield  {author} {\bibinfo {author} {\bibfnamefont {M.~J.}\ \bibnamefont
  {Holland}}\ and\ \bibinfo {author} {\bibfnamefont {K.}~\bibnamefont
  {Burnett}},\ }\href@noop {} {\bibfield  {journal} {\bibinfo  {journal}
  {Physical review letters}\ }\textbf {\bibinfo {volume} {71}},\ \bibinfo
  {pages} {1355} (\bibinfo {year} {1993})}\BibitemShut {NoStop}%
\bibitem [{\citenamefont {Datta}\ \emph {et~al.}(2011)\citenamefont {Datta},
  \citenamefont {Zhang}, \citenamefont {Thomas-Peter}, \citenamefont {Dorner},
  \citenamefont {Smith},\ and\ \citenamefont {Walmsley}}]{datta2011quantum}%
  \BibitemOpen
  \bibfield  {author} {\bibinfo {author} {\bibfnamefont {A.}~\bibnamefont
  {Datta}}, \bibinfo {author} {\bibfnamefont {L.}~\bibnamefont {Zhang}},
  \bibinfo {author} {\bibfnamefont {N.}~\bibnamefont {Thomas-Peter}}, \bibinfo
  {author} {\bibfnamefont {U.}~\bibnamefont {Dorner}}, \bibinfo {author}
  {\bibfnamefont {B.~J.}\ \bibnamefont {Smith}}, \ and\ \bibinfo {author}
  {\bibfnamefont {I.~A.}\ \bibnamefont {Walmsley}},\ }\href@noop {} {\bibfield
  {journal} {\bibinfo  {journal} {Physical Review A}\ }\textbf {\bibinfo
  {volume} {83}},\ \bibinfo {pages} {063836} (\bibinfo {year}
  {2011})}\BibitemShut {NoStop}%
\bibitem [{\citenamefont {{Wolfram Research, Inc.}}(2021)}]{cosecant}%
  \BibitemOpen
  \bibfield  {author} {\bibinfo {author} {\bibnamefont {{Wolfram Research,
  Inc.}}},\ }\href@noop {} {\enquote {\bibinfo {title} {Cosecant finite
  summation},}\ } (\bibinfo {year} {2021}),\ \bibinfo {note}
  {\url{http://functions.wolfram.com/01.10.23.0004.01}}\BibitemShut {NoStop}%
\bibitem [{\citenamefont {Guanzon}\ \emph {et~al.}(2020)\citenamefont
  {Guanzon}, \citenamefont {Lund},\ and\ \citenamefont
  {Ralph}}]{guanzon2020controllable}%
  \BibitemOpen
  \bibfield  {author} {\bibinfo {author} {\bibfnamefont {J.~J.}\ \bibnamefont
  {Guanzon}}, \bibinfo {author} {\bibfnamefont {A.~P.}\ \bibnamefont {Lund}}, \
  and\ \bibinfo {author} {\bibfnamefont {T.~C.}\ \bibnamefont {Ralph}},\
  }\href@noop {} {\bibfield  {journal} {\bibinfo  {journal} {Physical Review
  A}\ }\textbf {\bibinfo {volume} {102}},\ \bibinfo {pages} {032606} (\bibinfo
  {year} {2020})}\BibitemShut {NoStop}%
\bibitem [{\citenamefont {Hedayat}\ \emph {et~al.}(1978)\citenamefont
  {Hedayat}, \citenamefont {Wallis} \emph {et~al.}}]{hedayat1978hadamard}%
  \BibitemOpen
  \bibfield  {author} {\bibinfo {author} {\bibfnamefont {A.}~\bibnamefont
  {Hedayat}}, \bibinfo {author} {\bibfnamefont {W.~D.}\ \bibnamefont {Wallis}},
   \emph {et~al.},\ }\href@noop {} {\bibfield  {journal} {\bibinfo  {journal}
  {Annals of Statistics}\ }\textbf {\bibinfo {volume} {6}},\ \bibinfo {pages}
  {1184} (\bibinfo {year} {1978})}\BibitemShut {NoStop}%
\bibitem [{\citenamefont {Motes}\ \emph {et~al.}(2014)\citenamefont {Motes},
  \citenamefont {Gilchrist}, \citenamefont {Dowling},\ and\ \citenamefont
  {Rohde}}]{motes2014scalable}%
  \BibitemOpen
  \bibfield  {author} {\bibinfo {author} {\bibfnamefont {K.~R.}\ \bibnamefont
  {Motes}}, \bibinfo {author} {\bibfnamefont {A.}~\bibnamefont {Gilchrist}},
  \bibinfo {author} {\bibfnamefont {J.~P.}\ \bibnamefont {Dowling}}, \ and\
  \bibinfo {author} {\bibfnamefont {P.~P.}\ \bibnamefont {Rohde}},\ }\href@noop
  {} {\bibfield  {journal} {\bibinfo  {journal} {Physical review letters}\
  }\textbf {\bibinfo {volume} {113}},\ \bibinfo {pages} {120501} (\bibinfo
  {year} {2014})}\BibitemShut {NoStop}%
\bibitem [{\citenamefont {Slussarenko}\ and\ \citenamefont
  {Pryde}(2019)}]{slussarenko2019photonic}%
  \BibitemOpen
  \bibfield  {author} {\bibinfo {author} {\bibfnamefont {S.}~\bibnamefont
  {Slussarenko}}\ and\ \bibinfo {author} {\bibfnamefont {G.~J.}\ \bibnamefont
  {Pryde}},\ }\href@noop {} {\bibfield  {journal} {\bibinfo  {journal} {Applied
  Physics Reviews}\ }\textbf {\bibinfo {volume} {6}},\ \bibinfo {pages}
  {041303} (\bibinfo {year} {2019})}\BibitemShut {NoStop}%
\bibitem [{\citenamefont {Arrazola}\ \emph {et~al.}(2021)\citenamefont
  {Arrazola}, \citenamefont {Bergholm}, \citenamefont {Br{\'a}dler},
  \citenamefont {Bromley}, \citenamefont {Collins}, \citenamefont {Dhand},
  \citenamefont {Fumagalli}, \citenamefont {Gerrits}, \citenamefont {Goussev},
  \citenamefont {Helt} \emph {et~al.}}]{arrazola2021quantum}%
  \BibitemOpen
  \bibfield  {author} {\bibinfo {author} {\bibfnamefont {J.}~\bibnamefont
  {Arrazola}}, \bibinfo {author} {\bibfnamefont {V.}~\bibnamefont {Bergholm}},
  \bibinfo {author} {\bibfnamefont {K.}~\bibnamefont {Br{\'a}dler}}, \bibinfo
  {author} {\bibfnamefont {T.}~\bibnamefont {Bromley}}, \bibinfo {author}
  {\bibfnamefont {M.}~\bibnamefont {Collins}}, \bibinfo {author} {\bibfnamefont
  {I.}~\bibnamefont {Dhand}}, \bibinfo {author} {\bibfnamefont
  {A.}~\bibnamefont {Fumagalli}}, \bibinfo {author} {\bibfnamefont
  {T.}~\bibnamefont {Gerrits}}, \bibinfo {author} {\bibfnamefont
  {A.}~\bibnamefont {Goussev}}, \bibinfo {author} {\bibfnamefont
  {L.}~\bibnamefont {Helt}},  \emph {et~al.},\ }\href@noop {} {\bibfield
  {journal} {\bibinfo  {journal} {Nature}\ }\textbf {\bibinfo {volume} {591}},\
  \bibinfo {pages} {54} (\bibinfo {year} {2021})}\BibitemShut {NoStop}%
\bibitem [{\citenamefont {Percus}(2012)}]{percus2012combinatorial}%
  \BibitemOpen
  \bibfield  {author} {\bibinfo {author} {\bibfnamefont {J.~K.}\ \bibnamefont
  {Percus}},\ }\href@noop {} {\emph {\bibinfo {title} {Combinatorial
  methods}}},\ Vol.~\bibinfo {volume} {4}\ (\bibinfo  {publisher} {Springer
  Science \& Business Media},\ \bibinfo {year} {2012})\BibitemShut {NoStop}%
\end{thebibliography}%

\appendix

\section{The Separable Interferometer - Combinatorics of $n$ Single Photons into $m$ Modes} \label{sec:sepcomb}

\begin{table}[b]
    \caption{\label{tab:table1}%
        The combinatorics and Fisher information associated with $n$ single photons injected into the $m$ ports of the separable circuit $Y_m^{(\text{sep})} \equiv \oplus^{m/2}_{j=1} Y_2$. By considering how many of these photons land together in one $Y_2$, the $\binom{m}{n}$ possible cases can be separated based on Fisher information. 
    }
    \begin{ruledtabular}
        \begin{tabular}{ c c c c c }
            No. of  & No. of                & No. of                & Fisher    & No. of    \\ 
            Photons & $Y_2|11\rangle$       & $Y_2|10\rangle$ or    & Info.     & Cases     \\
            $n$     & $x$                   & $Y_2|01\rangle$       & $\mathcal{F}^{(\text{sep})}$       &           \\
            \hline
            1 & 0 & 1 & 1 & $\binom{m/2}{0}\binom{m/2-0}{1}2^1$ \\ 
            \hline
            2 & 0 & 2 & 2 & $\binom{m/2}{0}\binom{m/2-0}{2}2^2$ \\ 
              & 1 & 0 & 4 & $\binom{m/2}{1}\binom{m/2-1}{0}2^0$ \\ 
            \hline 
            3 & 0 & 3 & 3 & $\binom{m/2}{0}\binom{m/2-0}{3}2^3$ \\ 
              & 1 & 1 & 5 & $\binom{m/2}{1}\binom{m/2-1}{1}2^1$ \\     
            \hline 
            4 & 0 & 4 & 4 & $\binom{m/2}{0}\binom{m/2-0}{4}2^4$ \\ 
              & 1 & 2 & 6 & $\binom{m/2}{1}\binom{m/2-1}{2}2^2$ \\   
              & 2 & 0 & 8 & $\binom{m/2}{2}\binom{m/2-2}{0}2^0$ \\ 
            \hline 
            5 & 0 & 5 & 5 & $\binom{m/2}{0}\binom{m/2-0}{5}2^5$ \\ 
              & 1 & 3 & 7 & $\binom{m/2}{1}\binom{m/2-1}{3}2^3$ \\   
              & 2 & 1 & 9 & $\binom{m/2}{2}\binom{m/2-2}{1}2^1$ \\ 
        \end{tabular}
    \end{ruledtabular}
\end{table}

We can count the number of ways that $n$ single photons can be injected into the separable system $Y_m^{(\text{sep})} \equiv \oplus^{m/2}_{j=1} Y_2$ with arbitrary $m$ sizes. The combinatorics are summarised for the first few $n$ values in Table~\ref{tab:table1}, where we are separating the $\binom{m}{n}$ possible cases based on how many of the $n$ photons land together as a pair $x$ into one of the separated systems. The last column of Table~\ref{tab:table1} shows how many of these cases exist. For example, the expression for four photons $n=4$, with one pair $x=1$, is determined by multiplying together the following factors 
\begin{enumerate}
  \item $\binom{m/2}{1}$ is the number of ways that the one $Y_2|11\rangle$ can occur in the total $\oplus_{j=1}^{m/2}Y_2$ system.
  \item $\binom{m/2-1}{2}$ is the number of ways that two $Y_2|10\rangle$ can occur, minus the one $Y_2$ taken in step 1 already. 
  \item $2^2$ accounts for the fact that $Y_2|01\rangle$ can occur instead of $Y_2|10\rangle$. 
\end{enumerate}
Hence the pattern generated in the last column implies the following
\begin{align}
    \binom{m}{n} = \sum_{x=0}^{n/2} \binom{m/2}{x} \binom{m/2-x}{n-2x} 2^{n-2x}, \label{eq:mcn}
\end{align}
which we can verify is always mathematically true, independent of the physical basis in which this expression was derived. Now, if we assume that the photon source treats all modes equally, where it has an equal chance of generating any of the $\binom{m}{n}$ cases, then we can say that the probability that $x$ pairs occur is  
\begin{align}
    \mathcal{P}(x) = \binom{m}{n}^{-1}\binom{m/2}{x} \binom{m/2-x}{n-2x} 2^{n-2x}, 
\end{align}
where clearly $\sum_{x=0}^{n/2} \mathcal{P}(x) = 1$ according to Eq.~\eqref{eq:mcn}.

\section{The Uniform Interferometer - Calculating the Unitary and Hermitian Operator} \label{sec:uniherm} 

We first need to calculate the overall unitary matrix and then determine the generating Hermitian operator which dictates the action of this interferometer. By definition, the QFT is given by
\begin{align}
    (F_m)_{j,k} \equiv \frac{1}{\sqrt{m}} e^{-\frac{2i\pi(j-1)(k-1)}{m}}. 
\end{align}
The phase shift is applied to the top half of modes in $Z_m$ as follows
\begin{align}
    (Z_m)_{j,k} \equiv   
    \begin{cases}
        0, & j \neq k, \\
        e^{\frac{i\phi}{2}}, & j = k \leq m/2, \\
        e^{-\frac{i\phi}{2}}, & j = k > m/2,
    \end{cases}
\end{align}
where we introduced a physically inconsequential global phase shift of $e^{-\frac{i\phi}{2}}$ to simplify the computation. Combining these two unitaries will give
\begin{align}
    (F_mZ_m)_{j,k} &= 
    \begin{cases}
        \frac{1}{\sqrt{m}}e^{-\frac{2i\pi(j-1)(k-1)}{m}+\frac{i\phi}{2}}, & k \leq m/2, \\
        \frac{1}{\sqrt{m}}e^{-\frac{2i\pi(j-1)(k-1)}{m}-\frac{i\phi}{2}}, & k > m/2.
    \end{cases}
\end{align}
Thus, the full conjugation of the phase shifts $Z_m$ with QFTs $F_m$ will give us 
\begin{align}
    (Y^{(\text{uni})}_m)_{j,k} &= \sum_{p=1}^{m} (F_m Z_m)_{j,p} (F_m^\dag)_{p,k} \nonumber \\ 
    &= \sum_{p=1}^{m/2} \frac{e^{-\frac{2i\pi(j-1)(p-1)}{m}+\frac{i\phi}{2}+\frac{2i\pi(p-1)(k-1)}{m}}}{m}  \nonumber \\
    &\quad + \sum_{p=m/2+1}^{m} \frac{e^{-\frac{2i\pi(j-1)(p-1)}{m}-\frac{i\phi}{2}+\frac{2i\pi(p-1)(k-1)}{m}}}{m} \nonumber \\
    &= \frac{e^{\frac{i\phi}{2}}}{m} \sum_{p=1}^{m/2} e^{\frac{2i\pi(k-j)(p-1)}{m}} \nonumber \\ 
    &\quad + \frac{e^{-\frac{i\phi}{2}}}{m} \sum_{p=m/2+1}^{m} e^{\frac{2i\pi(k-j)(p-1)}{m}} \nonumber \\ 
    &= \frac{e^{\frac{i\phi}{2}}+ e^{-\frac{i\phi}{2}} e^{i\pi(k-j)}}{m} \sum_{p=1}^{m/2} e^{\frac{2i\pi(k-j)(p-1)}{m}} \nonumber \\ 
    &= \begin{cases}
        \cos(\phi/2), & j = k, \\
        \tfrac{e^{\frac{i\phi}{2}}+ e^{-\frac{i\phi}{2}} e^{i\pi(k-j)}}{m} \tfrac{1-e^{i\pi(k-j)}}{1-e^{\frac{2i\pi(k-j)}{m}}}, & j \neq k. 
    \end{cases}
\end{align}
We can simplify this further by looking for patterns in the $j$th row and $k$th column elements. In particular, if $k-j$ is an even number other than $0$ then $e^{i\pi(k-j)}=1$, however if $k-j$ is odd then $e^{i\pi(k-j)}=-1$. This means the scattering matrix which describes the overall uniform interferometer can be written as
\begin{align}
    (Y^{(\text{uni})}_m)_{j,k} &= \begin{cases}
        \cos(\phi/2), & j = k, \\
        0, & k - j \in \text{even}/\{0\}, \\
        \frac{ 4i \sin(\phi/2) }{m\left(1-e^{\frac{2i\pi(k-j)}{m}}\right)}, & k - j \in \text{odd}. 
    \end{cases}  
\end{align}
Using the matrix logarithm, the associated generator matrix is given by
\begin{align}
    (H^{(\text{uni})}_m)_{j,k} &= -\frac{i}{\phi}(\ln[Y^{(\text{uni})}_m])_{j,k}, \nonumber \\ 
    &= \begin{cases}
        0, & k - j \in \text{even}, \\
        \frac{2}{m\left(1-e^{\frac{2i\pi(k-j)}{m}}\right)}, & k - j \in \text{odd},
    \end{cases}
\end{align}
and the corresponding Hermitian operator is 
\begin{align}
    \hat{H}^{(\text{uni})} = \frac{2}{m} \sum^{m/2}_{j=1} \sum^{m/2}_{k=1} \left[ \tfrac{a^\dagger_{2j-1}a_{2k}}{1-e^{\frac{2i\pi(2k-2j+1)}{m}}} + \tfrac{a^\dagger_{2k}a_{2j-1}}{1-e^{-\frac{2i\pi(2k-2j+1)}{m}}} \right]. 
\end{align}

\section{The Symmetric Interferometer - Optical Circuit Decomposition for $T_4$ Symmetrising Transformation in Four Modes} \label{sec:symdecomp}  

\begin{figure}[htbp]
    \begin{center}
        \includegraphics[width=\linewidth]{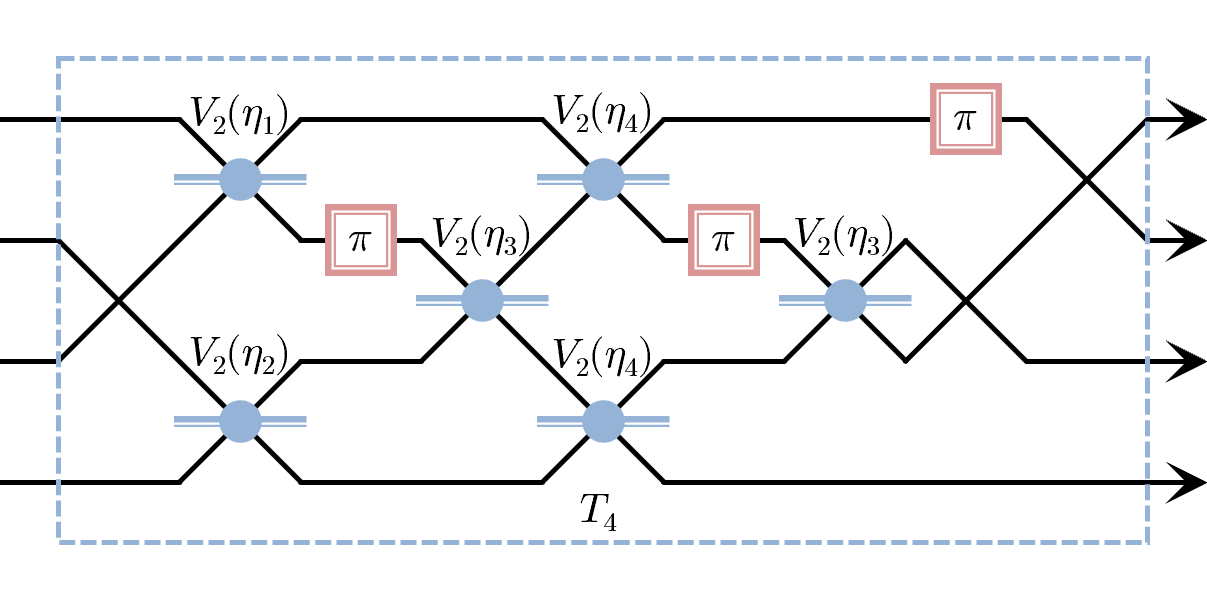}
        \caption{\label{fig:p3} 
            One possible decomposition of the four mode symmetrising transformation. The blue $V_2$ beam-splitters have a reflectivity of
            $\eta_1=1/\sqrt{50+20\sqrt{6}}\approx0.10$,
            $\eta_2=\sqrt{2/5}\approx0.63$,
            $\eta_3=1/\sqrt{251-100\sqrt{6}}\approx0.41$, and 
            $\eta_4=1/\sqrt{300+120\sqrt{6}}\approx0.04$, 
            while the red $\pi$ phase shifters induces a relative $-1$ phase difference in the indicated modes. 
            }
    \end{center}
\end{figure}

In Fig.~\ref{fig:p3} we give one possible decomposition, into base optical elements, of the $m=4$ mode symmetrising transformation $T_4$ given by 
\begin{align}
    T_4 = 
        \begin{bmatrix}
             0 & \tfrac{1}{\sqrt{2}} & \tfrac{1}{\sqrt{6}} & \tfrac{1}{\sqrt{3}} \\ 
             -\tfrac{1}{2} + \tfrac{1}{\sqrt{6}} & -\tfrac{1}{2\sqrt{3}} & \tfrac{1}{2} + \tfrac{1}{\sqrt{6}} & -\tfrac{1}{2\sqrt{3}} \\
            -\tfrac{1}{\sqrt{6}} & -\tfrac{1}{\sqrt{3}} & 0 & \tfrac{1}{\sqrt{2}} \\ 
            -\tfrac{1}{2} - \tfrac{1}{\sqrt{6}} & \tfrac{1}{2\sqrt{3}} & -\tfrac{1}{2} + \tfrac{1}{\sqrt{6}} & -\tfrac{1}{2\sqrt{3}} 
        \end{bmatrix},
\end{align}
as defined for general $m$ in Eq.~\eqref{eq:Tm}. This particular decomposition is based upon Clements et al. 2016 paper on the rectangular design of multiport interferometers \cite{clements2016optimal}. The blue two-mode beam-splitters are defined as 
\begin{align}
    V_2(\eta_i) \equiv \begin{bmatrix}
            \eta_i & -\sqrt{1-\eta_i^2} \\ 
            \sqrt{1-\eta_i^2} & \eta_i  
    \end{bmatrix},
\end{align}
where $\eta_i$ is the beam-splitter's particular reflectivity $\eta_i\in[0,1], \forall i$; the actual values for this decomposition are given in the caption of Fig.~\ref{fig:p3}. The red single-mode $\pi$ phase shift elements applies a $\exp(i \pi) = -1$ relative phase to the indicated modes.

\section{The Symmetric Interferometer - Comparison of QFI using Other Interferometers} \label{sec:symavg}

In this appendix we prove that the \emph{average} information gained from the separable $\langle \mathcal{F}^{(\text{sep})} \rangle$ and uniform $\langle \mathcal{F}^{(\text{uni})} \rangle$ interferometers are equivalent to the information gained from the symmetric interferometer $\mathcal{F}^{(\text{sym})}$. To do this, we will perform a calculation similar in spirit to the two photon example given in Table~\ref{tab:table2}, where the average is calculated directly by summing up the QFI from all possible cases and dividing by the total number of cases. Note that the scattershot source has an equal chance of generating $|n_1 n_2 n_3 \cdots n_m\rangle$ or $|n_2 n_1 n_3 \cdots n_m\rangle$, or any of the $m!$ different possible permutations. Hence, all we need to do is add up the QFI from all the $m!$ cases and divide the total by $m!$ (i.e. multiply by the probability $1/m!$) to get the average QFI. We will simplify the following calculations by just looking at the QFI excess above SNL.

\subsection{Average QFI using the Separable Interferometer}

From Eq.~\eqref{eq:sepF}, the QFI excess above SNL provided by the separable interferometer is
\begin{align}
    \mathcal{F}^{(\text{sep})} - n &= 2\sum_{j=1}^{m/2} n_{2j-1}n_{2j}. 
\end{align}
For now, we will focus on counting the information contribution just from the first $Y_2$ sub-experiment in the separable interferometer. Firstly, consider all the cases where $n_1$ is in the first mode and $n_2$ is in the second mode of the separable interferometer. These cases contributes an QFI excess of at least 
\begin{align}
    (m-2)! \times 2 n_1 n_2,  
\end{align}
where $(m-2)!$ is due to all the other permutations in the remaining $m-2$ modes. 
Next, consider all the cases where $n_1$ is in the first mode, then the contributing QFI from all these cases are 
\begin{align}
    2(m-2)! \sum_{k=2}^m n_1 n_k. 
\end{align}
Finally, consider all possible permutations, the first $Y_2$ must have contributed a QFI excess of 
\begin{align}
    2(m-2)! \sum_{j=1}^m \sum_{k\neq j}^m n_j n_k. 
\end{align}
However, there are $m/2$ of these $Y_2$'s in the separable interferometer, hence the total QFI excess from all $m!$ possible permutations must be
\begin{align}
    m(m-2)! \sum_{j=1}^m \sum_{k\neq j}^m n_j n_k. 
\end{align}
Now, we can get the average by dividing this result by the total number of case $m!$ as follows 
\begin{align}
    \langle \mathcal{F}^{(\text{sep})} \rangle - n &= \frac{1}{m-1} \sum_{j=1}^m \sum_{k\neq j}^m n_j n_k, \\ 
    \Rightarrow \langle \mathcal{F}^{(\text{sep})} \rangle &= \mathcal{F}^{(\text{sym})}.
\end{align}
We have hence shown that the QFI from the average scattershot sample in the separable interferometer is equal to the QFI from the symmetric interferometer. 

\subsection{Average QFI using the Uniform Interferometer}

We will now show the same for the uniform interferometer, where according to Eq.~\eqref{eq:uniF} the QFI excess above SNL is
\begin{align}
     \mathcal{F}^{(\text{uni})} - n &= \frac{8}{m^2} \sum_{j=1}^{m/2} \sum_{k=1}^{m/2} \frac{n_{2j-1}n_{2k}}{\sin^2\left(\frac{\pi(2k-2j+1)}{m}\right)}.
\end{align}
Now, we consider all the cases where $n_1$ is in the first mode and $n_2$ is in the second mode, these cases contribute a QFI excess of at least
\begin{align}
    (m-2)! \frac{8}{m^2} n_1n_2 \frac{1}{\sin^2\left(\frac{\pi}{m}\right)}, 
\end{align}
where $(m-2)!$ factor accounts for the possible permutations in the other $m-2$ modes. Now, consider the cases where $n_1$ is in any odd mode and $n_2$ is in any even mode, these cases add up to a QFI excess of 
\begin{align}
    (m-2)! &\frac{8}{m^2} n_1n_2 \sum_{j=1}^{m/2}\sum_{k=1}^{m/2}\frac{1}{\sin^2\left(\frac{\pi(2k-2j+1)}{m}\right)} \nonumber \\ &= m(m-2)! n_1 n_2. 
\end{align}
Note that we simplified the summation via the following $\sum_{j=1}^{m/2}\sum_{k=1}^{m/2}1/\sin^2\left(\frac{\pi(2k-2j+1)}{m}\right) = \sum_{j=1}^{m/2} \frac{m^2}{4} = \frac{m^3}{8}$, in which the first equality is due to a cosecant summation identity~\cite{cosecant}. Now, the previous expression represents all the QFI contribution of $n_1$ in the odd modes and $n_2$ in the even modes. Hence to get the total QFI contribution we sum over all possible cases as follows
\begin{align}
    m(m-2)! \sum_{j=1}^m \sum_{k\neq j}^m n_j n_k.  
\end{align}
Finally, by dividing this expression by the total number of cases $m!$ we can get the average QFI received using the uniform interferometer 
\begin{align}
    \langle \mathcal{F}^{(\text{uni})} \rangle - n &= \frac{1}{m-1} \sum_{j=1}^m \sum_{k\neq j}^m n_j n_k, \\ 
    \Rightarrow \langle \mathcal{F}^{(\text{uni})} \rangle &= \mathcal{F}^{(\text{sym})}.
\end{align}
Hence we have just shown that the QFI gained from the average scattershot sample using the uniform interferometer is equivalent to the QFI gained from any scattershot sample using the symmetric interferometer.

\section{The Symmetric Interferometer - Multiphoton Mode Invariance in Four Modes} \label{sec:syminv}

\begin{figure*}[htbp]
    \begin{center}
        \includegraphics[width=\linewidth]{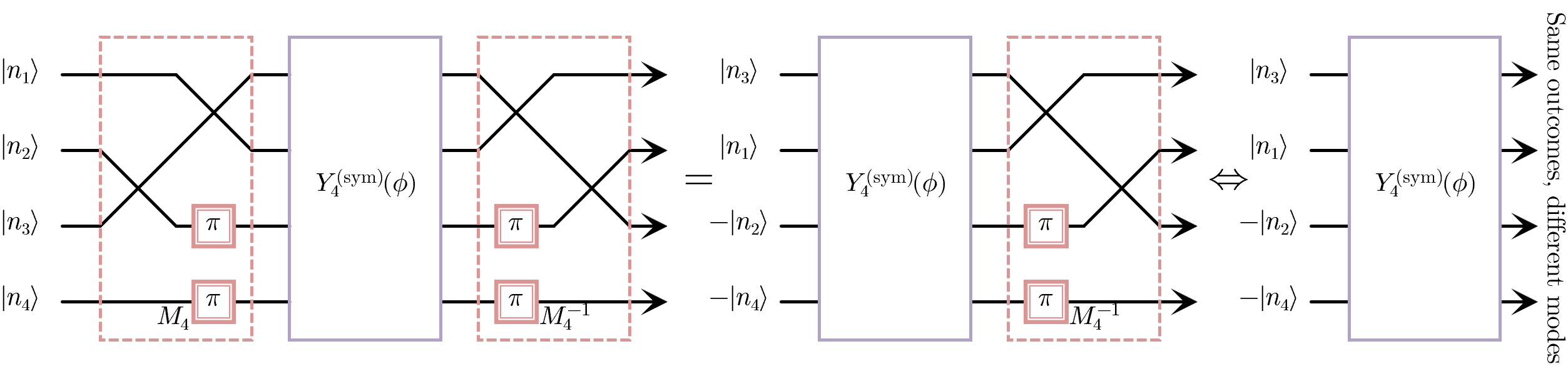}
        \caption{\label{fig:p4} 
            An example of how a particular mode permutation $M_4$ can be used to make input invariant statements. At the first step, the arbitrary input state is simply permuted $M_4^{-1}Y_4^{(\text{sym})}M_4|n_1n_2n_3n_4\rangle=M_4^{-1}Y_4^{(\text{sym})}|n_3n_1n_2n_4\rangle$. Then, it must be recognised that $M_4^{-1}Y_4^{(\text{sym})}|n_3n_1n_2n_4\rangle$ and $Y_4^{(\text{sym})} |n_3n_1n_2n_4\rangle$ have the same number measurement outcomes but at different modes, which means they have the same Fisher information as it's a sum over all outcomes. If there is a self-similarity transformation $Y_4^{(\text{sym})}=M_4^{-1}Y_4^{(\text{sym})}M_4$, then it follows that $Y_4^{(\text{sym})} |n_1n_2n_3n_4\rangle$ and $Y_4^{(\text{sym})} |n_3n_1n_2n_4\rangle$ also have the same Fisher information.} 
    \end{center}
\end{figure*}

We will prove here that our four mode metrology experimental apparatus, represented by 
\begin{align}
    Y_4^{(\text{sym})}(\phi) &= 
        \begin{bmatrix}
             c & -s &  s &  s \\ 
             s &  c &  s & -s \\
            -s & -s &  c & -s \\ 
            -s &  s &  s &  c 
        \end{bmatrix}, \\ 
        c &= \cos (\phi/2),\quad s(m=4) = \frac{\sin (\phi/2)}{\sqrt{3}}, 
\end{align}
will give the same Fisher information irrespective of which input ports the photons are injected into, using completely arbitrary photon states $|n_1n_2n_3n_4\rangle, n_i\in\mathbb{N}$. 

First, to get an idea of how this is going to be done, consider the following example transformation
\begin{align}
    M_4 &= 
        \begin{bmatrix}
             0 &  1 &  0 &  0 \\ 
             0 &  0 & -1 &  0 \\
             1 &  0 &  0 &  0 \\ 
             0 &  0 &  0 & -1 
        \end{bmatrix},
\end{align}
which permutes the modes $(2,3,1)$. The negative phase factors at certain modes means that the following self-similarity transformation 
\begin{align}
    M_4^{-1} Y_4^{(\text{sym})}(\phi) M_4 = Y_4^{(\text{sym})}(\phi) \label{eq:selfsim}
\end{align}
is true, which we can show by direct computation. This is a very useful relationship, as it allows us to make statements about the input invariance as summarised in Fig.~\ref{fig:p4}. In other words, we can show that 
\begin{align}
    Y_4^{(\text{sym})}(\phi)|n_1n_2n_3n_4\rangle &= M_4^{-1}Y_4^{(\text{sym})}(\phi)M_4|n_1n_2n_3n_4\rangle, \nonumber \\ 
    &= M_4^{-1}Y_4^{(\text{sym})}(\phi)|n_3n_1n_2n_4\rangle, \nonumber \\ 
    &\Leftrightarrow Y_4^{(\text{sym})}(\phi)|n_3n_1n_2n_4\rangle. 
\end{align}
The last step just means that the number measurement outcomes from $Y_4^{(\text{sym})}|n_3n_1n_2n_4\rangle$ will be the same as $M_4^{-1}Y_4^{(\text{sym})}|n_3n_1n_2n_4\rangle$, however which output port these outcomes occur in is different due to $M_4^{-1}$. Since the Fisher information can be calculated by the summation over all outcomes, we can conclude that the Fisher information is the same in both cases $Y_4^{(\text{sym})}|n_3n_1n_2n_4\rangle \Leftrightarrow M_4^{-1}Y_4^{(\text{sym})}|n_3n_1n_2n_4\rangle$. Finally, due to the self-similarity transformation, we can ultimately conclude that $Y_4^{(\text{sym})}|n_3n_1n_2n_4\rangle \Leftrightarrow Y_4^{(\text{sym})}|n_1n_2n_3n_4\rangle$; in other words, the Fisher information with an input of $|n_1n_2n_3n_4\rangle$ will be the same as $|n_3n_1n_2n_4\rangle$.

If we find more self-similarity transformations like $M_4$, we can make similar input invariance statements for all possible permutations of the modes. In fact, we only really need to show self-similarity using the following two permutations 
\begin{align}
    M_{4,1} &= 
        \begin{bmatrix}
             0 &  1 &  0 &  0 \\ 
             1 &  0 &  0 &  0 \\
             0 &  0 & -1 &  0 \\ 
             0 &  0 &  0 &  1 
        \end{bmatrix},\quad  
    M_{4,2} = 
        \begin{bmatrix}
             0 &  1 &  0 &  0 \\ 
             0 &  0 &  1 &  0 \\
             0 &  0 &  0 &  1 \\ 
            -1 &  0 &  0 &  0 
        \end{bmatrix}. 
\end{align}
The first one switches the first two modes $(1,2)$, while the second is a permutation over all modes $(2,3,4,1)$; these two transformations are sufficient to generate all other possible permutations. We can show by direct computation that  
\begin{align}
    M_{4,a}^{-1} Y_4^{(\text{sym})}(\phi) M_{4,a} = [Y_4^{(\text{sym})}(\phi)]^T = Y_4^{(\text{sym})}(-\phi), \nonumber
\end{align}
for $a \in \{1,2\}$ where these similarity transformations gives the transpose of $Y_4^{(\text{sym})}(\phi)$, rather than itself. However, this is effectively the same as the self-similarity transformation in Eq.~\eqref{eq:selfsim} due to the property $[Y_m^{(\text{sym})}(\phi)]^T = Y_m^{(\text{sym})}(-\phi)$ (as the off-diagonals are $\propto \sin(\phi/2)$), and since the measurement procedure we are conducting is near $\phi \approx 0$. Hence we can use the same logic as in Fig.~\ref{fig:p4} with not only with $M_{4,1}$ and $M_{4,2}$, but with all multiplicative combinations $M_{4,x}=\prod_j M_{4,a_j}$ of these two transformations, which can produce all possible mode permutations. Hence we know that the Fisher information with an input of $|n_1n_2n_3n_4\rangle$ will be the same even if an arbitrary permutation was performed $M_{4,x}|n_1n_2n_3n_4\rangle$ for any $x$, therefore we have proven which input ports of $Y_4^{(\text{sym})}$ the photons are injected into is metrologically irrelevant.

\section{The Symmetric Interferometer - Determining Outcome Probabilities for Single Photon Inputs} \label{sec:symprob}

The probability that the input is equal to the output is given by 
\begin{align}
    P_= = |p_=|^2,\quad p_= = \langle \vec{n} | Y_m^{(\text{sym})} | \vec{n} \rangle = \text{Per}(\mathcal{Y}_n), 
\end{align}
where $\mathcal{Y}_n=(Y_m^{(\text{sym})})_{j,k}$ is an $n \times n$ submatrix of $Y_m^{(\text{sym})}$, composed of all the $j$th row and $k$th column elements in which $n_j=n_k=1$, and Per is the matrix permanent function. As a concrete example, consider the following four mode metrology experiment 
\begin{align}
    Y_4^{(\text{sym})} &= 
        \begin{bmatrix}
             c & -s &  s &  s \\ 
             s &  c &  s & -s \\
            -s & -s &  c & -s \\ 
            -s &  s &  s &  c 
        \end{bmatrix}. 
\end{align}
The probability amplitude for the input $|\vec{n}\rangle=|1110\rangle$ is the permanent of 
\begin{align}
    \mathcal{Y}_3(1110) = (Y_4^{(\text{sym})})_{j\in\{1,2,3\},k\in\{1,2,3\}} = 
        \begin{bmatrix}
             c & -s &  s \\ 
             s &  c &  s \\
            -s & -s &  c 
        \end{bmatrix}, \nonumber
\end{align}
however the probability amplitude for the input $|\vec{n}\rangle=|1011\rangle$ is the permanent of 
\begin{align}
    \mathcal{Y}_3(1011) = (Y_4^{(\text{sym})})_{j\in\{1,3,4\},k\in\{1,3,4\}} = 
        \begin{bmatrix}
             c &  s &  s \\ 
            -s &  c & -s \\ 
            -s &  s &  c 
        \end{bmatrix}. \nonumber
\end{align}
Hence, the probability amplitude associated with a single photon input $|\vec{n}\rangle$ is always related to the permanents of the following matrices 
\begin{align}
    \mathcal{Y}_n &= 
       \overbrace{ \begin{bmatrix}
                c & -r_1 s & -r_2 s & \cdots \\ 
            r_1 s &      c & -r_3 s & \cdots \\
            r_2 s &  r_3 s &      c & \cdots \\ 
            \vdots & \vdots & \vdots & \ddots 
        \end{bmatrix} }^{n \text{ columns}} 
        \left. \phantom{\begin{bmatrix}
                 c \\ 
            r_1 s \\
            r_2 s \\ 
            \vdots 
        \end{bmatrix}\hspace{-3em}} \right\} \rotatebox[origin=c]{-90}{\scriptsize $n$ rows}, \nonumber \\ 
        &= \begin{bmatrix}
                0 & -r_1 s & -r_2 s & \cdots \\ 
            r_1 s &      0 & -r_3 s & \cdots \\
            r_2 s &  r_3 s &      0 & \cdots \\ 
            \vdots & \vdots & \vdots & \ddots 
        \end{bmatrix} +
        \begin{bmatrix}
            c & 0 & 0 & \cdots \\ 
            0 & c & 0 & \cdots \\
            0 & 0 & c & \cdots \\ 
            \vdots & \vdots & \vdots & \ddots 
        \end{bmatrix}, \nonumber \\ 
        &= S_n + C_n, \quad r_j \in \{-1,1\},
\end{align}
where we abstracted the location of the off-diagonal negatives via $r_j$. This abstraction allows us to make general statements irrespective of where the photons have entered $Y_m$. Note that $C_n = \cos(\phi/2) \mathbb{I}_n$ contains the diagonal elements, while $S_n$ contains the off-diagonal elements in which the skew-symmetric property holds $S_n = -S_n^T$. 

It is known that given two $n \times n$ square matrices $A_n$ and $B_n$, we can expand the permanent as follows 
\begin{align}
    \text{Per}(A_n + B_n) = \sum_{x,y} \text{Per}(A_n)_{j \in x, k \in y} \text{Per}(B_n)_{j \in \bar{x}, k \in \bar{y}}, \nonumber
\end{align}
where $x$ and $y$ are same sized subsets of $\{1,\cdots,n\}$, while $\bar{x}$ and $\bar{y}$ are the complementary subsets~\cite{percus2012combinatorial}. However, since $C_n$ is proportional to the identity, we actually have a simplified equation in our case
\begin{align}
    \text{Per}(\mathcal{Y}_n) &= \text{Per}(S_n + C_n), \nonumber \\
    &= \sum_{x} \text{Per}(S_n)_{j \in x, k \in x} \text{Per}(C_n)_{j \in \bar{x}, k \in \bar{x}}, 
\end{align}
where there is just one sum over all the possible subsets $x\subseteq\{1,\cdots,n\}$, and $\bar{x}=\{1,\cdots,n\} \backslash x$ is the complement set. We will show that this permanent expansion equation leads to a straightforward pattern as we increase $n$. 

Firstly, consider the expansion of the permanent for $n=2$, which corresponds to taking the permanent of the submatrices $x\in\{\{\ \},\{1\},\{2\},\{1,2\}\}$ in $S_2$, and multiplying it with the permanent of the complementary submatrices in $C_2$, as follows
\begin{align}
    \text{Per}(\mathcal{Y}_2) &= \text{Per}(S_2 + C_2), \nonumber \\
    &= \text{Per}\left( 
        \begin{bmatrix}
                 0 & -r_1 s \\ 
            r_1 s &       0 
        \end{bmatrix} +
        \begin{bmatrix}
            c & 0 \\ 
            0 & c 
        \end{bmatrix}
    \right), \nonumber \\ 
    &= \text{Per}[\ ]\text{Per}\begin{bmatrix}
            c & 0 \\ 
            0 & c 
        \end{bmatrix}
        + \binom{2}{1}\text{Per}[0]\text{Per}[c] \nonumber \\
        &\quad + \text{Per}\begin{bmatrix}
                 0 & -r_1 s \\ 
            r_1 s &       0 
        \end{bmatrix}\text{Per}[\ ], \nonumber \\ 
    &= c^2 - s^2.
\end{align}
where we define $\text{Per}[\ ]=1$.
We note from above that $\text{Per}(S_2)=-s^2$, however since $r_j^2=1$ for all sign placeholders, we can state generally 
\begin{align}
    \text{Per}\begin{bmatrix}
                 0 & -r_j s \\ 
            r_j s &       0 
        \end{bmatrix} = -s^2,\quad \forall j. 
\end{align}
It is also apparent that the only contribution of $C_n$ to each term will only be 
\begin{align}
    \text{Per}(C_d) = \text{Per}(c\mathbb{I}_d) = c^d, 
\end{align}
where $d$ is the size of the matrix at each term. 

Next, consider the expansion for $n=3$ photons, which likewise is associated with the submatrices $x\in\{\{\ \},\{1\},\{2\},\{3\},\{1,2\},\{1,3\},\{2,3\},\{1,2,3\}\}$ in $S_3$, and the complementary submatrices in $C_3$, as follows
\begin{align}
    \text{Per}(\mathcal{Y}_3) &= \text{Per}(S_3 + C_3), \nonumber \\
    &= \text{Per}\left( 
        \begin{bmatrix}
                0 & -r_1 s & -r_2 s \\ 
            r_1 s &      0 & -r_3 s \\
            r_2 s &  r_3 s &      0 
        \end{bmatrix} +
        \begin{bmatrix}
            c & 0 & 0 \\ 
            0 & c & 0 \\
            0 & 0 & c 
        \end{bmatrix}
    \right) \nonumber \\ 
    &= \text{Per}[\ ]\text{Per}\begin{bmatrix}
            c & 0 & 0 \\ 
            0 & c & 0 \\
            0 & 0 & c 
        \end{bmatrix}
        + \binom{3}{1}\text{Per}[0]\text{Per}\begin{bmatrix}
            c & 0 \\ 
            0 & c  
        \end{bmatrix} \nonumber \\
        &\quad + \binom{3}{2}\text{Per}\begin{bmatrix}
                 0 & -r_i s \\ 
            r_i s &       0 
        \end{bmatrix}\text{Per}[c] \nonumber \\
        &\quad + \text{Per}\begin{bmatrix}
                0 & -r_1 s & -r_2 s \\ 
            r_1 s &      0 & -r_3 s \\
            r_2 s &  r_3 s &      0 
        \end{bmatrix}\text{Per}[\ ], \nonumber \\
    &= c^3 - \binom{3}{2}cs^2.
\end{align}
It is obvious that $\text{Per}(S_1)=\text{Per}[0]=0$, and one can calculate directly that $\text{Per}(S_3)=0$, however a more general statement can be made if we consider
\begin{align}
    \text{Per}(S_d) = \text{Per}(-S_d^T) = (-1)^d\text{Per}(S_d),
\end{align}
in which we are taking advantage of the skew-symmetry property and invariance of the permanent under transposition. Hence this means that if $d$ is odd then 
\begin{align}
    \text{Per}(S_d) &= -\text{Per}(S_d) \nonumber \\
    \Rightarrow \text{Per}(S_d) &= 0,\quad \forall\text{ odd } d. 
\end{align}
This means we can ignore all terms which contribute to odd powers of $s$ in our expansion. 

If we continue to expand to higher orders of $n$, we can get the following equations
\begin{align}
    \text{Per}(\mathcal{Y}_4) &= c^4 - \binom{4}{2}c^2s^2 + k_{4,4}s^4, \\ 
    \text{Per}(\mathcal{Y}_5) &= c^5 - \binom{5}{2}c^3s^2 + k_{5,4}cs^4, \\ 
    \text{Per}(\mathcal{Y}_6) &= c^6 - \binom{6}{2}c^4s^2 + k_{6,4}c^2s^4 + k_{6,6}s^6,
\end{align}
where these $k_{n,d}$ coefficients actually depend on the particular off-diagonal signs $r_j$ contained within the matrix. Based on this permanent expansion technique, and permanents of $S_d$ and $C_d$, it is clear that for arbitrary $n$ we have the following polynomial expansion 
\begin{align}
    p_= &= \text{Per}(\mathcal{Y}_n) = \sum_{j=0}^{\lfloor n/2 \rfloor} k_{n,2j} c^{n-2j} s^{2j} \nonumber \\
    &= c^n - \binom{n}{2} c^{n-2} s^2 + \sum_{j=2}^{\lfloor n/2 \rfloor} k_{n,2j} c^{n-2j} s^{2j}. 
\end{align}
We can use this probability amplitude expression to calculate the QFI, as it is not necessary with certain measurement schemes to know all the terms.

\begin{figure*}[htbp]
    \begin{center}
        \includegraphics[width=\linewidth]{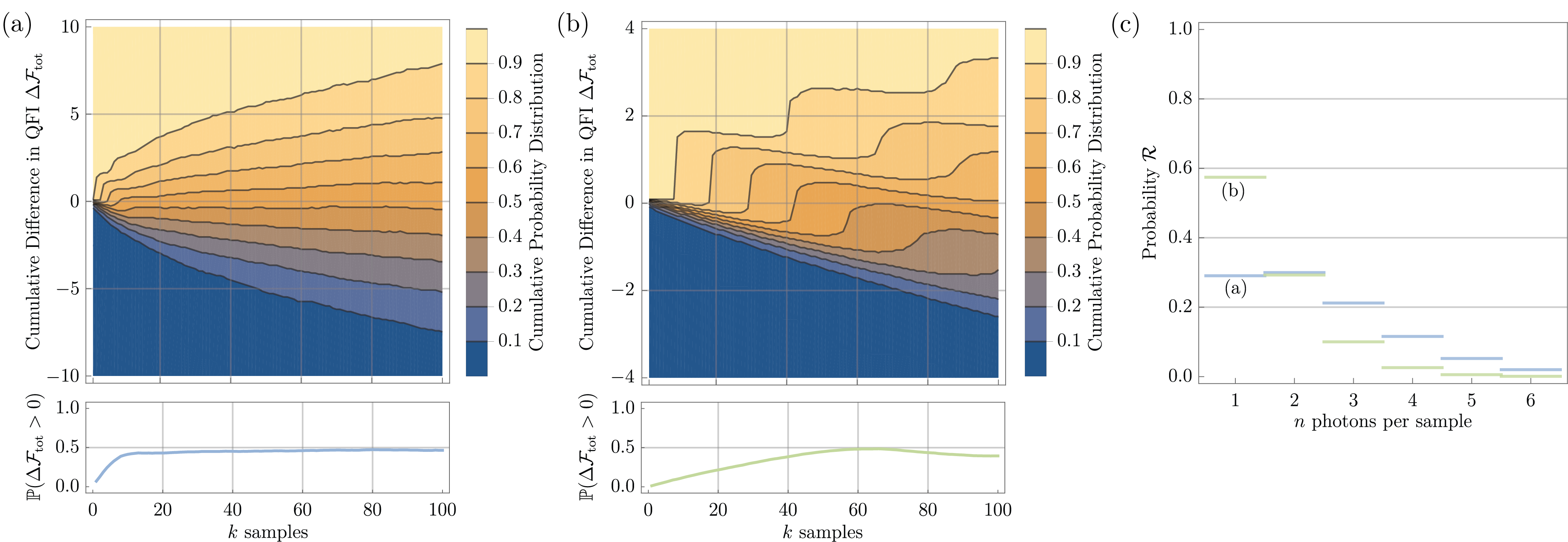}
        \caption{\label{fig:MC2} 
            (a) Similar to Fig.~\ref{fig:MC}, where the upper graph is the cumulative probability distribution of $10^4$ random walkers that individually track the running total QFI difference $\Delta \mathcal{F}_{\text{tot}}(k) = \mathcal{F}^{(\text{sep})}_{\text{tot}}(k) - \mathcal{F}^{(\text{sym})}_{\text{tot}}(k)$ as more random samples are taken. The lower graph is the probability that the separable experiment gave more information $\mathbb{P}[\mathcal{F}^{(\text{sep})}_{\text{tot}}(k) > \mathcal{F}^{(\text{sym})}_{\text{tot}}(k)]$, for a given sample size $k$. This was done using a small system with $m=2^{5}$ modes and random samples from a scattershot source with $\chi = 0.25$ squeezers.
            (b) The same graphs as (a), however with more modes $m=2^{6}$ and weaker squeezing $\chi = 0.125$. 
            (c) The photon statistics of the samples.
            } 
    \end{center}
\end{figure*} 

\section{Monte Carlo Simulation using Bunched Photons and Low Modes} \label{sec:advanbunch}

We perform the same Monte Carlo simulation as described in Section~\ref{sec:advan}, however allowing for multiple photons per mode $n_j \in \mathbb{N}$ for the random scattershot inputs. We are also considering smaller systems, where Fig.~\ref{fig:MC2}(a) was run with the parameters $(m=2^5,\chi=0.25)$ and Fig.~\ref{fig:MC2}(b) was run with $(m=2^6,\chi=0.125)$. We can see in Fig.~\ref{fig:MC2}(b) that weaker squeezing results in the oscillating features appearing, however not as prominently as in Fig.~\ref{fig:MC}. This reasoning is reinforced by the Fock basis measurement statistics of the samples $\mathcal{R}$ given in Fig.~\ref{fig:MC2}(c), which shows that for the (b) parameter set single photons are the most likely outcome. This figure also shows how decreasing the number of photons, or increasing the number of modes, will increase the sample size advantage of the symmetric interferometer as shown in the lower $\mathbb{P}(\Delta \mathcal{F}_{\text{tot}}(k)>0)$ graphs. Based on this numerical analysis, the overall features due to scaling the parameters are expected to be the same as the single photon situation described in Section~\ref{sec:advan}.

\begin{figure*}[htbp]
    \begin{center}
        \includegraphics[width=\linewidth]{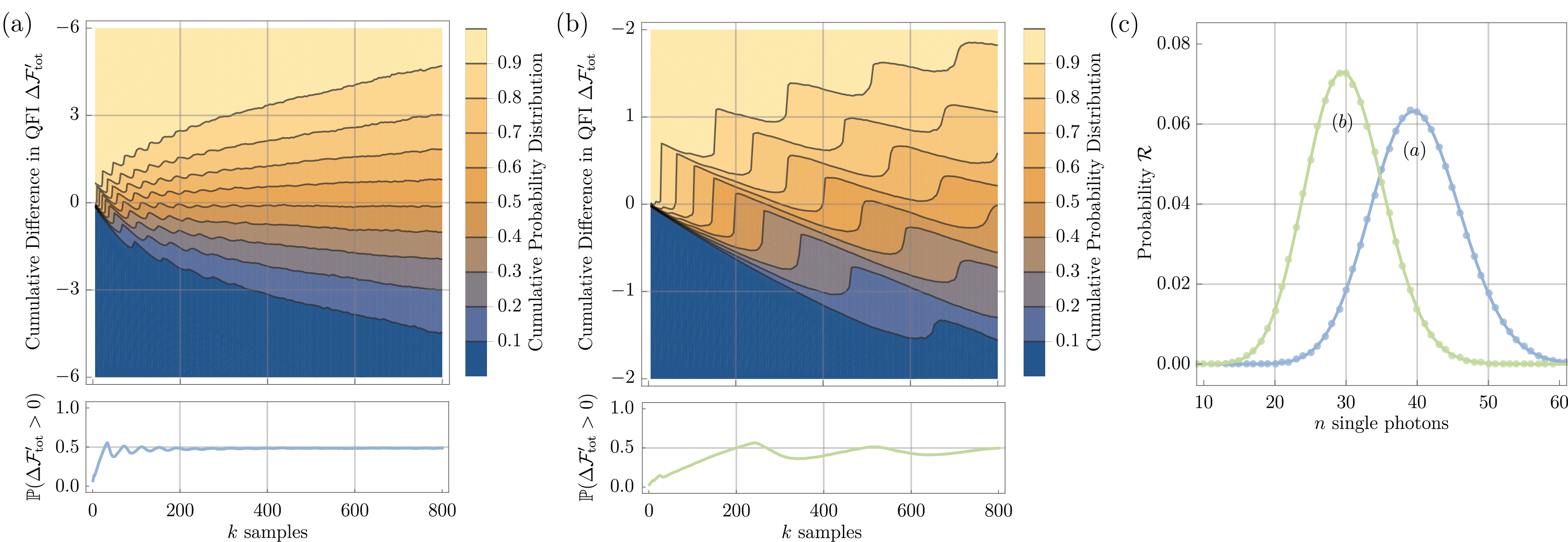}
        \caption{\label{fig:MC3} 
            (a) Similar to Fig.~\ref{fig:MC}, where the upper graph is the cumulative probability distribution of $10^4$ random walkers that individually track the running total QFI difference $\Delta \mathcal{F}'_{\text{tot}}(k) = \mathcal{F}^{(\text{uni})}_{\text{tot}}(k) - \mathcal{F}^{(\text{sym})}_{\text{tot}}(k)$ as more random samples are taken. The lower graph is the probability that the uniform experiment gave more information $\mathbb{P}[\mathcal{F}^{(\text{uni})}_{\text{tot}}(k) > \mathcal{F}^{(\text{sym})}_{\text{tot}}(k)]$, for a given sample size $k$. This was done using a system with $m=2^{16}$ modes and samples from a scattershot source with $\chi\approx 0.0247$ squeezers.
            (b) The same graphs as (a), however with more modes $m=2^{18}$ and weaker squeezing $\chi\approx 0.0107$. 
            (c) The photon statistics of the samples, which shows that these particular $m$ and $\chi$ parameters translates to an average of $\langle n \rangle = 40$ photons for (a) and $\langle n \rangle = 30$ photons for (b). 
            } 
    \end{center}
\end{figure*} 

\section{Monte Carlo Simulation Comparing the Uniform and Symmetric Interferometers} \label{sec:advanuni}

We perform the same Monte Carlo simulation as described in Section~\ref{sec:advan}, however we are instead comparing the uniform interferometer with the symmetric interferometer as summarised in Fig.~\ref{fig:MC3}. Note that this is with multiple single photons and large mode sizes.

\end{document}